\documentclass[
    prd,
    twocolumn,
    nopacs,
    floatfix,
    amsmath,
    nofootinbib,
    amssymb,
    preprintnumbers
]{revtex4-2}

\usepackage{graphicx,subfigure,color,dcolumn,booktabs}
\usepackage{array}
\usepackage{tabularx}
\usepackage{txfonts}
\usepackage{slashed}
\usepackage{epstopdf}
\usepackage{amsmath}
\usepackage{appendix}
\usepackage{bm}

\usepackage[a4paper,margin=2.0cm]{geometry}
\usepackage{amsmath,amssymb,bm}
\usepackage{xcolor}
\usepackage{tikz}
\usepackage{pgfplots}
\usepackage{booktabs}
\usepackage{adjustbox}
\usepackage{siunitx}
\usepackage{float}

\usetikzlibrary{arrows.meta}
\pgfplotsset{compat=1.18}

\begin{document}
\newcommand{\pdlr}{\overset{\leftrightarrow}{\partial}}
\newcommand{\vect}[1]{\boldsymbol{#1}}

\title{Bound-state spectra of $\chi_{cJ}$ in finite nuclei and the universal pattern of mass levels}

\author{Tian-Le Gao$^{1,2,4}$}\email{gaotl2025@lzu.edu.cn}
\author{Ze-Hua Zhang$^{1,5,6}$}
\author{Xiang Liu$^{1,2,3,4}$}\email{xiangliu@lzu.edu.cn}
\affiliation{
	$^1$School of Physical Science and Technology, Lanzhou University, Lanzhou 730000, China\\
	$^2$Lanzhou Center for Theoretical Physics, Key Laboratory of Theoretical Physics of Gansu Province, Key Laboratory of Quantum Theory and Applications of MoE and Gansu Provincial Research Center for Basic Disciplines of Quantum Physics, Lanzhou University, Lanzhou 730000, China\\
	$^3$MoE Frontiers Science Center for Rare Isotopes, Lanzhou University, Lanzhou 730000, China\\
	$^4$Research Center for Hadron and CSR Physics, Lanzhou University and Institute of Modern Physics of CAS, Lanzhou 730000, China\\
    $^5$Key Laboratory of Nuclear Physics and Ion-beam Application~(MOE), Institute of Modern Physics, Fudan University, Shanghai $200433$, China\\
    $^6$Shanghai Research Center for Theoretical Nuclear Physics, NSFC and Fudan University, Shanghai 200438, China
}

\begin{abstract}
In this work, we investigate possible $\chi_{cJ}$--nuclear bound states with $J=0,1,2$ using in-medium mass shifts generated by virtual $D^{(*)}\bar{D}^{(*)}$ loops within an unquenched framework. The resulting $\chi_{cJ}$--nucleus potentials are constructed in the local density approximation, and the bound state spectra are calculated for $^{12}{\rm C}$, $^{16}{\rm O}$, $^{40}{\rm Ca}$, $^{90}{\rm Zr}$, $^{197}{\rm Au}$, and $^{208}{\rm Pb}$. Bound states are obtained for all systems considered. The $\chi_{c0}(1P)$ and $\chi_{c1}(1P)$ spectra are nearly degenerate, whereas the larger in-medium mass shift of $\chi_{c2}(1P)$ leads to deeper binding. Although the absolute bound state energies depend appreciably on the cutoff parameter, the energy differences relative to the $1s$ level are considerably less sensitive to it and exhibit a regular pattern that decreases approximately as $A^{-2/3}$ with increasing nuclear mass number. A cosh-type potential with a common nuclear geometry provides a compact description of these spectra. The predicted bound-state structures and level-spacing systematics could be investigated in future high-statistics near-threshold photoproduction experiments at the upgraded JLab facility.
\end{abstract}

\maketitle

\section{Introduction}
\label{sec:introduction}

Heavy quarkonia offer a unique window into the nonperturbative regime of strong interaction. Although the vacuum spectrum of low-lying charmonium states is successfully described by quenched potential models~\cite{Eichten:1979ms,Godfrey:1985dt,Voloshin:2007dx,Brambilla:2010cs}, it has long been recognized that coupled-channel effects, particularly those involving virtual open-charm meson loops, provide crucial corrections, especially in the vicinity of thresholds~\cite{Ono:1983rd,vanBeveren:2003kd,Kalashnikova:2005ui,Li:2009ad,Ortega:2009hj,Danilkin:2010cc} (see the review article \cite{Bai:2026atm} and references therein for more details). The behavior of these hidden-flavor states in a nuclear environment has consequently garnered significant interest, spanning a wide range of phenomena: from the possibility of charmonium-nucleus bound states~\cite{Brodsky:1989jd,Wasson:1991fb,Kaidalov:1992hd,Belyaev:2006vn,Tsushima:2011kh,Yokota:2013sfa,Yokota:2014lma}, to in-medium spectral modifications, the properties of exotic candidates~\cite{ExHIC:2010gcb,ExHIC:2011say,Cho:2013rpa,Abreu:2016qci,ExHIC:2017smd,Azizi:2017ubq,Zhang:2020dwn,Albaladejo:2021cxj,Montesinos:2023qbx,Montesinos:2024uhq}, and the long-standing puzzle of charmonium production and suppression in heavy-ion collisions~\cite{Matsui:1986dk,Rapp:2008tf,Cobos-Martinez:2017vtr,Tang:2025ypa}.

The formation of charmonium-nucleus bound states represents a particularly direct and incisive probe of the interaction between hidden-charm matter and ordinary nuclear matter. This system is theoretically pristine: since charmonia and nucleons share no light valence quarks, the dominant light-meson exchange mechanisms are suppressed by the Okubo-Zweig-Iizuka rule. Consequently, the interaction is governed by more subtle, fundamental QCD processes, such as multi-gluon exchange, QCD color van der Waals forces, and virtual open-charm meson loops~\cite{Brodsky:1989jd,Luke:1992tm,Kharzeev:1994pz}. This has motivated a rich body of theoretical work employing diverse tools, including the QCD multipole expansion~\cite{Luke:1992tm}, QCD sum rules~\cite{Klingl:1998sr,Hayashigaki:1998ey,Kim:2000kj}, effective Lagrangians and quark-based nuclear approaches~\cite{Krein:2010vp,Tsushima:2011kh}, phenomenological potentials~\cite{Yokota:2013sfa}, and lattice QCD~\cite{Kawanai:2010ev,Beane:2014sda}. Despite these efforts, predictions for the bound state energies of even the lightest $\eta_c$ and $J/\psi$ mesons in finite nuclei remain highly model-dependent~\cite{Tsushima:2011kh,Cobos-Martinez:2020ynh,Zeminiani:2021vaq,Kaur:2026qzf}, underscoring the need for a robust, microscopic framework that can systematically connect the internal structure of the charmonium to the nuclear medium.

An effective framework for this connection is the quark-meson coupling (QMC) model, which self-consistently links the nuclear mean fields to the internal quark structure of hadrons~\cite{Guichon:1987jp,Guichon:1995ue}. In this model, light-quark mean fields modify the properties of hadrons that contain light valence quarks. For conventional charmonia, however, the heavy quarks do not couple directly to these fields, implying no direct mass shift at the valence level. Their in-medium modification instead arises through a more subtle and fascinating mechanism: the virtual dissociation into $D^{(*)}\bar D^{(*)}$ pairs. The open-charm mesons in these intermediate states \emph{do} contain light valence quarks, making their effective masses density dependent. The resulting charmonium loop self-energies thus become a sensitive probe of the nuclear environment, encoding the influence of the medium through the internal dynamics of the coupled-channel system~\cite{Krein:2010vp}.

The $P$-wave triplet $\chi_{cJ}(1P)$ ($J=0,1,2$) presents an ideal testing ground for this mechanism. The distinct spin structures of its members lead to qualitatively different couplings to the $D\bar D$, $D\bar D^*+\mathrm{c.c.}$, and $D^*\bar D^*$ channels. Crucially, the $\chi_{c2}(1P)$ couples to the $D^*\bar D^*$ channel in an $S$-wave, rendering its self-energy exceptionally sensitive to the vector-vector loop and its medium modifications~\cite{Zhang:2025fol}. This channel dependence makes the $\chi_{cJ}$ triplet an unparalleled laboratory for disentangling the role of different open-charm mesons in nuclear binding. Furthermore, this system is phenomenologically vital, as the radiative decays of $\chi_{c1}(1P)$ and $\chi_{c2}(1P)$ significantly contribute to the observed $J/\psi$ yield in heavy-ion collisions. Understanding their cold nuclear matter modifications is therefore not just an academic exercise but a prerequisite for reliably interpreting charmonium suppression data and discriminating between cold and hot nuclear matter effects~\cite{Sibirtsev:1999jr,Golubeva:2002jz}.

While a recent QMC-based analysis has successfully calculated the density-dependent mass shifts of the $\chi_{cJ}(1P)$ triplet in uniform nuclear matter, the crucial next step---connecting these results to the observable realm of finite nuclei---has yet to be performed within this consistently microscopic framework. To bridge this gap, we map the density-dependent mass shifts onto radial $\chi_{cJ}$-nucleus potentials using the local-density approximation and solve the Klein-Gordon equation to determine the bound-state spectra of these charmonia in nuclei from $^{12}$C to $^{208}$Pb. This work, for the first time within the QMC framework, provides a comprehensive finite-nucleus analysis of the entire $\chi_{cJ}(1P)$ triplet. We systematically investigate the dependence of the bound state energies on the charmonium spin, nuclear size, and the cutoff parameter $\alpha$, and present a detailed analysis of the level ordering and excitation energies. To facilitate future applications and provide a robust benchmark, we introduce a phenomenological cosh-type parametrization that faithfully captures the common radial geometry of the calculated potentials. Our results not only shed light on the intricate interplay between hidden charm and nuclear matter but also establish a concrete, theoretically grounded reference for future experimental searches and theoretical studies of charmonium-nucleus bound states.

The remainder of this paper is organized as follows. In Sec.~II, we summarize the QMC description, the in-medium open-charm-loop mechanism, the construction of the finite-nucleus potentials, and the bound-state equation. Section~III presents the calculated potentials and spectra and discusses their dependence on the charmonium state, nuclear mass number, and cutoff parameter, with particular emphasis on the relative level spacings. In Sec.~IV, we introduce the cosh-type parametrization and benchmark it against the reference potentials and spectra. Finally, Sec.~V contains a summary and concluding remarks.

\section{THEORETICAL FRAMEWORK}\label{sec2}
In this section, we outline the theoretical framework used to construct $\chi_{cJ}$--nucleus bound states. We first introduce the QMC description of nuclear matter and finite nuclei, in which the light quarks in the open-charm mesons $D^{(*)}$ and $\bar D^{(*)}$ respond to the nuclear scalar and vector mean fields. Since the $\chi_{cJ}(1P)$ states are hidden-charm $c\bar c$ mesons, their medium modifications are generated indirectly through the density-dependent $D^{(*)}\bar D^{(*)}$ loop self-energies. The resulting mass shift is then converted, within the local density approximation, into a real $\chi_{cJ}$--nucleus potential. Finally, the bound state energies are obtained by solving a Klein--Gordon-type equation.

For cold symmetric nuclear matter, the QMC Lagrangian density at the hadronic level is written as~\cite{Guichon:1995ue}
\begin{equation}
\begin{aligned}
\mathcal{L}_{\rm QMC}^{0}
=&\,\bar{\psi}_N
\left[
i\gamma\cdot\partial
-M_N^*(\hat{\sigma})
-g_\omega \hat{\omega}_\mu\gamma^\mu
\right]\psi_N
\\
&+\frac{1}{2}
\left(
\partial_\mu\hat{\sigma}\,\partial^\mu\hat{\sigma}
-m_\sigma^2\hat{\sigma}^2
\right)
\\
&-\frac{1}{2}
\left[
\partial_\mu\hat{\omega}_\nu
\left(
\partial^\mu\hat{\omega}^\nu-\partial^\nu\hat{\omega}^\mu
\right)
-m_\omega^2\hat{\omega}_\mu\hat{\omega}^\mu
\right].
\end{aligned}
\label{eq:qmc_lagrangian}
\end{equation}
Here $\psi_N$ is the nucleon field, and $\hat{\sigma}$ and $\hat{\omega}^\mu$ denote the scalar-isoscalar and vector-isoscalar meson fields, respectively. In the mean-field approximation, these meson fields are replaced by their ground-state expectation values. For uniform nuclear matter at rest, translational and rotational invariance lead to $\hat{\sigma}\to \sigma$ and $\hat{\omega}^\mu\to \delta^{\mu 0}\omega$. The superscript ``0'' indicates that the isovector $\rho$-meson field and the Coulomb interaction are omitted in symmetric nuclear matter, but must be included for finite nuclei with unequal proton and neutron numbers~\cite{Guichon:1995ue,Saito:1996yb,Tsushima:1998ru,Cobos-Martinez:2023hbp,Mondal:2023iwe,Zeminiani:2024dyo,Mondal:2024vyt,Mondal:2025qxm}.

The scalar field modifies the internal structure of the nucleon through the field-dependent effective mass
\begin{equation}
M_N^*(\sigma)=M_N-g_\sigma(\sigma)\sigma ,
\label{eq:nucleon_mass}
\end{equation}
where $g_\sigma(\sigma)$ is the effective scalar coupling at the hadronic level. In the QMC model, this field dependence originates from the response of the confined light quarks inside the nucleon, which is described using the MIT bag model.

For a hadron $h$ described as an MIT bag, the light constituents 
$q=u,d$ couple directly to the scalar and vector mean fields. In the rest frame
of nuclear matter, their in-medium Dirac equations are~\cite{Tsushima:1997df,Tsushima:2002cc}
\begin{equation}
\left[i\gamma\cdot\partial-(m_q-g_\sigma^q\sigma)\mp \gamma^0 g_\omega^q\omega\right]
\begin{pmatrix}\psi_q\\ \psi_{\bar q}\end{pmatrix}=0,
\label{eq:light_quark_bag}
\end{equation}
where the upper and lower signs correspond to a light quark and antiquark, respectively. The constants $g_\sigma^q$ and $g_\omega^q$ are the quark-level couplings to the $\sigma$ and $\omega$ mean fields. In uniform nuclear matter, they are related to the corresponding nucleon-level couplings by
\begin{align}
   \frac{\partial}{\partial\sigma}
\left[g_\sigma(\sigma)\sigma\right]
=&\,3g_\sigma^q S_N(\sigma),
\\
g_\omega=&\,3g_\omega^q , 
\end{align}
where $S_N(\sigma)$ is the scalar density of a light quark in the nucleon bag,
given by
\begin{equation}
S_N(\sigma)=\frac{\Omega_q^*/2+m_q^* R_N^* \left(\Omega_q^*-1\right)}{\Omega_q^*\left(\Omega_q^*-1\right)+m_q^* R_N^*/2}.
\label{eq:SN_definition}
\end{equation}
Thus, the field dependence of $g_\sigma(\sigma)$ reflects the response of the internal nucleon structure to the external scalar field.

Heavy constituent quarks, denoted by $Q$, are assumed not to couple directly to these light-meson mean fields,
\begin{equation}
\left(i\gamma\cdot\partial-m_Q\right)\psi_{Q,\bar Q}=0.
\label{eq:heavy_quark_bag}
\end{equation}

The corresponding energy of an in-medium (anti)quark in a hadron can be expressed as
\begin{align}
\begin{pmatrix}\epsilon_q^*\\ \epsilon_{\bar q}^*\end{pmatrix}&=\Omega_q^* \pm R_h^*V_\omega^q, \\
\epsilon_Q^* &= \epsilon_{\bar Q}^* = \Omega_Q^* .
\end{align}
For any constituent flavor $f$ in the hadron $h$,
\begin{equation}
\Omega_f^*(\sigma,R_h^*)=\left[x_f^{*2}+\left(R_h^*m_f^*\right)^2\right]^{1/2},
\label{eq:Omega_definition}
\end{equation}
where $m_q^*=m_{\bar q}^*=m_q-g_\sigma^q\sigma$ and $m_Q^*=m_{\bar Q}^*=m_Q$. The lowest-mode bag eigenfrequency $x_f^*$ is determined by the bag boundary
condition
\begin{equation}
j_0(x_f^*)=
\sqrt{\frac{\Omega_f^*-m_f^*R_h^*}{\Omega_f^*+m_f^*R_h^*}}\,j_1(x_f^*) .
\label{eq:bag_boundary_condition}
\end{equation}
Here, $j_0$ and $j_1$ are spherical Bessel functions.

The in-medium bag mass of the hadron is then obtained from
\begin{align} 
&m_h^*(\sigma)=
\frac{\sum_f n_f\Omega_f^*(\sigma,R_h^*)-z_h}{R_h^*}
+\frac{4\pi}{3}(R_h^*)^3B,\label{eq:bag_mass}\\
&\left.\frac{\partial m_h^*(\sigma,R)}{\partial R}\right|_{R=R_h^*}=0.
\label{eq:bag_mass2}
\end{align}
Here, $n_f$ is the number of constituents of flavor $f$ in $h$, $B$ is the bag constant, and $z_h$ parametrizes center-of-mass and gluonic corrections which can be fixed in vacuum by reproducing the physical hadron mass together with the stability condition~\cite{Guichon:1995ue}.

The formalism described above can be applied directly to hadrons containing light valence constituents. In particular, the open-charm mesons $D^{(*)}$ and $\bar D^{(*)}$ contain one light antiquark or light quark, respectively, and their density-dependent masses in symmetric nuclear matter can therefore be calculated within the QMC framework. The numerical results for these masses can be found in Refs.~\cite{Tsushima:1998ru,Tsushima:2011kh}.

\begin{figure}[htbp]
\centering
\begin{tikzpicture}[
    line/.style={line width=1.1pt},
    vertex/.style={circle, fill=black, inner sep=1.7pt},
    label/.style={font=\large}
]

\def\R{1.15}
\def\xL{-3.0}
\def\xR{3.0}

\draw[line] (\xL,0) -- (-\R,0);
\draw[line] (\R,0) -- (\xR,0);

\draw[line] (0,0) circle (\R);

\node[vertex] at (-\R,0) {};
\node[vertex] at (\R,0) {};

\node[label] at (-2.25,0.45) {$\chi_{cJ}$};
\node[label] at (2.25,0.45) {$\chi_{cJ}$};

\node[label] at (0,1.52) {$D^{(*)}$};
\node[label] at (0,-1.52) {$\bar{D}^{(*)}$};

\end{tikzpicture}
\caption{
Schematic illustration of the virtual $D^{(*)}\bar{D}^{(*)}$ loop contribution to the self-energy of the $\chi_{cJ}(1P)$ state in nuclear matter.
}
\label{fig:chicJ-loop}
\end{figure}
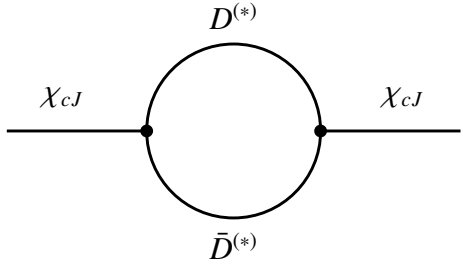

The $\chi_{cJ}(1P)$ states, on the other hand, are hidden-charm $c\bar c$ mesons and contain no light valence quarks. Thus, they do not couple directly to the nuclear mean fields at the valence-quark level in the present QMC description. Their in-medium mass shifts are instead evaluated in an unquenched picture~\cite{Tsushima:2011kh,Zeminiani:2020aho,Zeminiani:2023gqc,Zhang:2025fol}. In this approach, the $\chi_{cJ}(1P)$ states couple virtually to intermediate $D^{(*)}\bar D^{(*)}$ channels (see Fig.~\ref{fig:chicJ-loop}), whose masses are modified by the nuclear medium. The density dependence of these open-charm intermediate states then induces a self-energy correction to the $\chi_{cJ}(1P)$ masses.

In the present work, an effective Lagrangian approach is used to
describe the loop contribution to the $\chi_{cJ}(1P)$ self-energy~\cite{Wise:1992hn,Burdman:1992gh,Casalbuoni:1996pg,Yan:1992gz}. The effective interaction Lagrangian for the coupling of the $1P$ charmonium states to $D^{(*)}\bar D^{(*)}$ pairs is written as~\cite{Huang:2021kfm,Colangelo:2003sa,Chen:2013cpa,Chen:2014ccr,Zhang:2025fol}
\begin{align}
\mathcal{L}_{\chi_{cJ}D^{(*)}D^{(*)}}
=&\,
-g_{\chi_{c0}DD}\,\chi_{c0}D\bar D
-g_{\chi_{c0}D^*D^*}\,\chi_{c0}\bar D_\mu^*D^{*\mu}
\nonumber\\
&+i g_{\chi_{c1}D^*D}\,\chi_{c1}^{\mu}
\left(D\bar D_\mu^*-D_\mu^*\bar D\right)
\nonumber\\
&-g_{\chi_{c2}DD}\,\chi_{c2}^{\mu\nu}
\partial_\nu\bar D\,\partial_\mu D
+g_{\chi_{c2}D^*D^*}\,\chi_{c2}^{\mu\nu}
\bar D_\mu^*D_\nu^*
\nonumber\\
&-i g_{\chi_{c2}D^*D}\,
\varepsilon_{\mu\nu\alpha\beta}
\partial^\alpha\chi_{c2}^{\mu\rho}
\left(
\partial^\beta\bar D\,\partial_\rho D^{*\nu}
-\partial_\rho\bar D^{*\nu}\,\partial^\beta D
\right).
\label{eq:chicj_lag}
\end{align}

For a given intermediate channel $l$, the corresponding self-energy
contribution can be expressed schematically as
\begin{equation}
\Pi_l(p^2;\rho_B)
=
i\int\frac{d^4q}{(2\pi)^4}
\frac{
N_l(p,q)\,
F_1(q^2,m_1^{2})\,
F_2((p-q)^2,m_2^{2})
}{
\left(q^2-m_1^{*2}+i\epsilon\right)
\left[(p-q)^2-m_2^{*2}+i\epsilon\right]
}.
\label{eq:generic_loop}
\end{equation}
Here $l$ labels the allowed $D^{(*)}\bar D^{(*)}$ loop channel,
$m_1^*$ and $m_2^*$ are the density-dependent QMC masses of the intermediate open-charm mesons, and $N_l(p,q)$ contains the vertex structure and spin projection factors. The explicit forms of $N_l(p,q)$ can be found in Ref.~\cite{Zhang:2025fol}. In evaluating these loop integrals, only the scalar effective masses of the intermediate $D^{(*)}$ mesons are retained, since the vector mean-field contributions cancel between the meson and anti-meson because of baryon-number conservation~\cite{Tsushima:2011kh}.

The loop integral is regularized by a dipole form factor at each vertex,
\begin{equation}
F_i(k_i^2,m_i^{2})
=
\left(
\frac{m_i^{2}-\Lambda_i^2}
     {k_i^2-\Lambda_i^2}
\right)^2,
\label{eq:dipole}
\end{equation}
where $k_i$ and $m_i$ are the four-momentum and mass of the
exchanged open-charm meson. The cutoff is parametrized as
\begin{equation}
\Lambda_i=m_i+\alpha\Lambda_{\rm QCD},
\label{eq:cutoff}
\end{equation}
with $\Lambda_{\rm QCD}=220~\mathrm{MeV}$ and $\alpha$ a dimensionless
parameter of order unity~\cite{Cheng:2004ru}.

The in-medium masses of the $\chi_{cJ}(1P)$ triplet are then determined
self-consistently from
\begin{align}
\left[M_{\chi_{cJ}}^*(\rho_B)\right]^2
=&\,
M_{\chi_{cJ}}^2
-\sum_l \mathrm{Re}\,\Pi_l(M_{\chi_{cJ}}^2;0)
\nonumber\\
&+\sum_l
\mathrm{Re}\,\Pi_l
\left(
\left[M_{\chi_{cJ}}^*(\rho_B)\right]^2;\rho_B
\right),
\label{eq:in_medium_mass}
\end{align}
where $J=0,1,2$ and $M_{\chi_{cJ}}$ denotes the physical vacuum mass. The
second term subtracts the vacuum loop contribution already included in the
physical mass, while the last term adds the loop correction evaluated with
density-dependent intermediate $D^{(*)}\bar D^{(*)}$ masses. The mass shift is
therefore defined as
\begin{equation}
\Delta M_{\chi_{cJ}}(\rho_B)
=
M_{\chi_{cJ}}^*(\rho_B)-M_{\chi_{cJ}} .
\label{eq:chicj_mass_shift}
\end{equation}

To apply the above nuclear-matter results to finite nuclei, we use the local density approximation. Namely, the meson embedded at a position $r$ inside a nucleus is assumed to experience the same in-medium modification as in uniform nuclear matter at the local baryon density $\rho_B(r)$~\cite{Tsushima:1998ru,Tsushima:2011kh}. Therefore, in the following we calculate the radial density distribution $\rho_B(r)$ of the finite nucleus and use it to obtain the spatial potential distribution $V_{\chi_{cJ}-A}(r)$.

For finite nuclei with unequal proton and neutron numbers, the isovector $\rho$ meson and the Coulomb field must be included in addition to the scalar $\sigma$ and vector $\omega$ fields. For a static and spherically symmetric nucleus, define
\begin{equation}
\nabla_r^2\equiv \frac{d^2}{dr^2}+\frac{2}{r}\frac{d}{dr},
\label{eq:radial_laplace}
\end{equation}
and write the occupied nucleon spinors in terms of radial functions $G_\alpha(r)$ and $F_\alpha(r)$, normalized as
\begin{equation}
\int_0^\infty dr\,
\left(|G_\alpha(r)|^2+|F_\alpha(r)|^2\right)=1.
\label{eq:nucleon_norm}
\end{equation}
With
\begin{equation}
d_\alpha(r)=\frac{2j_\alpha+1}{4\pi r^2},
\label{eq:d_alpha}
\end{equation}
the scalar, baryon, isovector, and proton densities are
\begin{align}
\rho_s(r)&=\sum_{\alpha}^{\rm occ}d_\alpha(r)
\left(|G_\alpha(r)|^2-|F_\alpha(r)|^2\right),
\label{eq:rho_s}\\
\rho_B(r)&=\sum_{\alpha}^{\rm occ}d_\alpha(r)
\left(|G_\alpha(r)|^2+|F_\alpha(r)|^2\right),
\label{eq:rho_b}\\
\rho_3(r)&=2\sum_{\alpha}^{\rm occ}d_\alpha(r)t_\alpha
\left(|G_\alpha(r)|^2+|F_\alpha(r)|^2\right),
\label{eq:rho_3}\\
\rho_p(r)&=\sum_{\alpha}^{\rm occ}d_\alpha(r)
\left(t_\alpha+\frac{1}{2}\right)
\left(|G_\alpha(r)|^2+|F_\alpha(r)|^2\right),
\label{eq:rho_p}
\end{align}
where $t_\alpha$ is the eigenvalue of $\tau_3^N/2$ and the proton convention is $t_\alpha=+1/2$.

The meson and Coulomb fields satisfy the self-consistent radial equations
\begin{align}
\left(\nabla_r^2-m_\sigma^2\right)\sigma(r)
&=-g_\sigma C_N[\sigma(r)]\rho_s(r),
\label{eq:sigma_finite}\\
\left(\nabla_r^2-m_\omega^2\right)\omega(r)
&=-g_\omega\rho_B(r),
\label{eq:omega_finite}\\
\left(\nabla_r^2-m_\rho^2\right)b(r)
&=-\frac{g_\rho}{2}\rho_3(r),
\label{eq:rho_finite}\\
\nabla_r^2 A(r)
&=-e\rho_p(r).
\label{eq:coulomb_finite}
\end{align}
Here, $b(r)$ denotes the time component of the neutral $\rho$-meson mean field and $e$ represents the unit charge. The factor $C_N(\sigma)$ describes the scalar response of the nucleon bag and satisfies $g_\sigma C_N(\sigma)=-\partial M_N^*(\sigma)/\partial\sigma$. In the QMC-I parametrization~\cite{Saito:2005rv}, the field-dependent scalar coupling in Eq.~(2) is $g_\sigma(\sigma)=g_\sigma[1-(a_N/2)g_\sigma\sigma]$, or equivalently $M_N^*(\sigma)=M_N-g_\sigma[1-(a_N/2)g_\sigma\sigma]\sigma$. It then follows that
\begin{equation}
g_\sigma C_N[\sigma(r)]=g_\sigma\left[1-a_N g_\sigma \sigma(r)\right].
\end{equation}

The corresponding radial Dirac equations for a nucleon state $\alpha$ are
\begin{align}
\frac{dG_\alpha}{dr}+\frac{\kappa_\alpha}{r}G_\alpha-\bigg[
\epsilon_\alpha-g_\omega\omega(r)-t_\alpha g_\rho b(r)&\nonumber\\
-\left(t_\alpha+\frac{1}{2}\right)eA(r)+M_N^*(r)
\bigg]F_\alpha&=0,
\label{eq:dirac_G}\\
\frac{dF_\alpha}{dr}-\frac{\kappa_\alpha}{r}F_\alpha+\bigg[
\epsilon_\alpha-g_\omega\omega(r)-t_\alpha g_\rho b(r)&\nonumber\\
-\left(t_\alpha+\frac{1}{2}\right)eA(r)-M_N^*(r)
\bigg]G_\alpha&=0,
\label{eq:dirac_F}
\end{align}
where $\epsilon_\alpha$ is the energy.

The corresponding total energy of the finite nucleus is written as
\begin{align}
	E_{\rm tot}
	=&\,
	\sum_{\alpha}^{\rm occ}
	(2j_\alpha+1)\epsilon_\alpha
	-\frac{1}{2}
	\int d^3r\,
	\bigg[
	-g_\sigma C_N(\sigma(r))\sigma(r)\rho_s(r)\nonumber\\
	&\,+g_\omega\omega(r)\rho_B(r)
	+\frac{g_\rho}{2}b(r)\rho_3(r)
	+eA(r)\rho_p(r)
	\bigg].
	\label{eq:finite_nucleus_total_energy}
\end{align}

Solving Eqs.~\eqref{eq:sigma_finite}--\eqref{eq:finite_nucleus_total_energy} gives the finite-nucleus density profile $\rho_B^A(r)$ used in the local density approximation.

Once the density-dependent mass shift $\Delta M_{\chi_{cJ}}(\rho_B;\alpha)$ has been obtained in uniform nuclear matter, the potential in a finite nucleus $A$ is constructed by the local density approximation,
\begin{equation}
V_{\chi_{cJ}-A}(r)
=
\Delta M_{\chi_{cJ}}\,\!\big[\rho_B^A(r)\big].
\label{eq:lda_potential}
\end{equation}

The bound state energies are calculated from the Klein-Gordon equation with the reduced mass of the $\chi_{cJ}$-nucleus system,
\begin{align}
\mu_A^{(J)}=
\frac{M_{\chi_{cJ}}M_A}{M_{\chi_{cJ}}+M_A},
\label{eq:reduced_mass}
\end{align}
where $M_A$ is the nuclear mass.  The wave equation is~\cite{Lu:1994wz,Kwan:1978zh}
\begin{equation}
\left[-\nabla^2+\big(\mu_A^{(J)}\big)^2+2\mu_A^{(J)}V_{\chi_{cJ}-A}(r)\right]
\phi_{n\ell m}^{(J,\,A)}(\bm r)
=\mathcal{E}_{n\ell}^{2}\,
\phi_{n\ell m}^{(J,\,A)}(\bm r).
\label{eq:kg}
\end{equation}
The term $[V_{\chi_{cJ}-A}(r)]^2$ is neglected, which is well
justified because the mass shift is much smaller than the reduced
mass of the $\chi_{cJ}$--nucleus system.

For a central potential,
\begin{equation}
\phi_{n\ell m}^{(J,\,A)}(\bm r)=\frac{u_{n\ell}^{(J,\,A)}(r)}{r}Y_{\ell m}(\hat{\bm r}).
\end{equation}
 Here, $u_{n\ell}^{(J,\,A)}(r)$ is the reduced radial wave function
describing the radial motion of the $\chi_{cJ}(1P)$ relative to the
nucleus, while $Y_{\ell m}(\hat{\mathbf r})$ is the spherical
harmonic describing the angular dependence.

The bound state energy is defined as
\begin{equation}
E_{n\ell}^{(J,\,A)}=\mathcal{E}_{n\ell}-\mu_A^{(J)},
\label{eq:binding_energy}
\end{equation}
which is negative for a bound state. For $\chi_{c0}(1P)$, Eq.~\eqref{eq:kg} is directly the scalar-meson equation. For $\chi_{c1}(1P)$ and $\chi_{c2}(1P)$, the same Klein--Gordon-type equation is used for the center-of-mass motion by neglecting polarization-dependent splittings~\cite{Tsushima:2011kh}.

\section{numerical results}
In the present calculations, we employ the same QMC-I parameter
set as in Refs.~\cite{Saito:2005rv,Zhang:2025fol}, with $a_N=8.8\times10^{-4}\ {\rm MeV}^{-1}$~\cite{Saito:2005rv}. In the previous work~\cite{Zhang:2025fol}, the in-medium mass shifts of the $\chi_{cJ}(1P)$ states in infinite nuclear matter were evaluated within the QMC model. For convenience of the following discussion, we list in Table~\ref{tab:mass-shift-chicj-rho0} their values at normal nuclear matter density, $\rho_0=0.15$ fm$^{-3}$, for different choices of the cutoff parameter $\alpha$~\cite{Zeminiani:2020aho,Qian:2023taw,Gao:2024qth,Zhang:2025fol}. The results show that the mass shifts of $\chi_{c0}(1P)$ and $\chi_{c1}(1P)$ are nearly identical for a given value of $\alpha$, whereas the $\chi_{c2}(1P)$ state exhibits a much larger downward mass shift. Moreover, the magnitude of the mass shift increases with increasing $\alpha$ for all three $\chi_{cJ}(1P)$ states.

\begin{table}[htbp]
    \centering
    \caption{Mass shifts $\Delta M_{\chi_{cJ}}$ of $\chi_{cJ}(1P)$ states at normal nuclear matter density $\rho_0=0.15$ fm$^{-3}$ for different cutoff parameters $\alpha$, in units of MeV.}
    \label{tab:mass-shift-chicj-rho0}
    \begin{tabular*}{0.95\columnwidth}{@{\extracolsep{\fill}}cccc}
        \toprule\toprule
        State & $\alpha=2$ & $\alpha=3$ & $\alpha=4$ \\
        \midrule
        $\chi_{c0}(1P)$ & $-30.76$ & $-43.93$ & $-53.47$ \\
        $\chi_{c1}(1P)$ & $-30.87$ & $-44.63$ & $-54.27$ \\
        $\chi_{c2}(1P)$ & $-59.36$ & $-81.00$ & $-94.59$ \\
        \bottomrule\bottomrule
    \end{tabular*}
\end{table}

\begin{figure}[htbp]
    \centering
    \includegraphics[width=8cm]{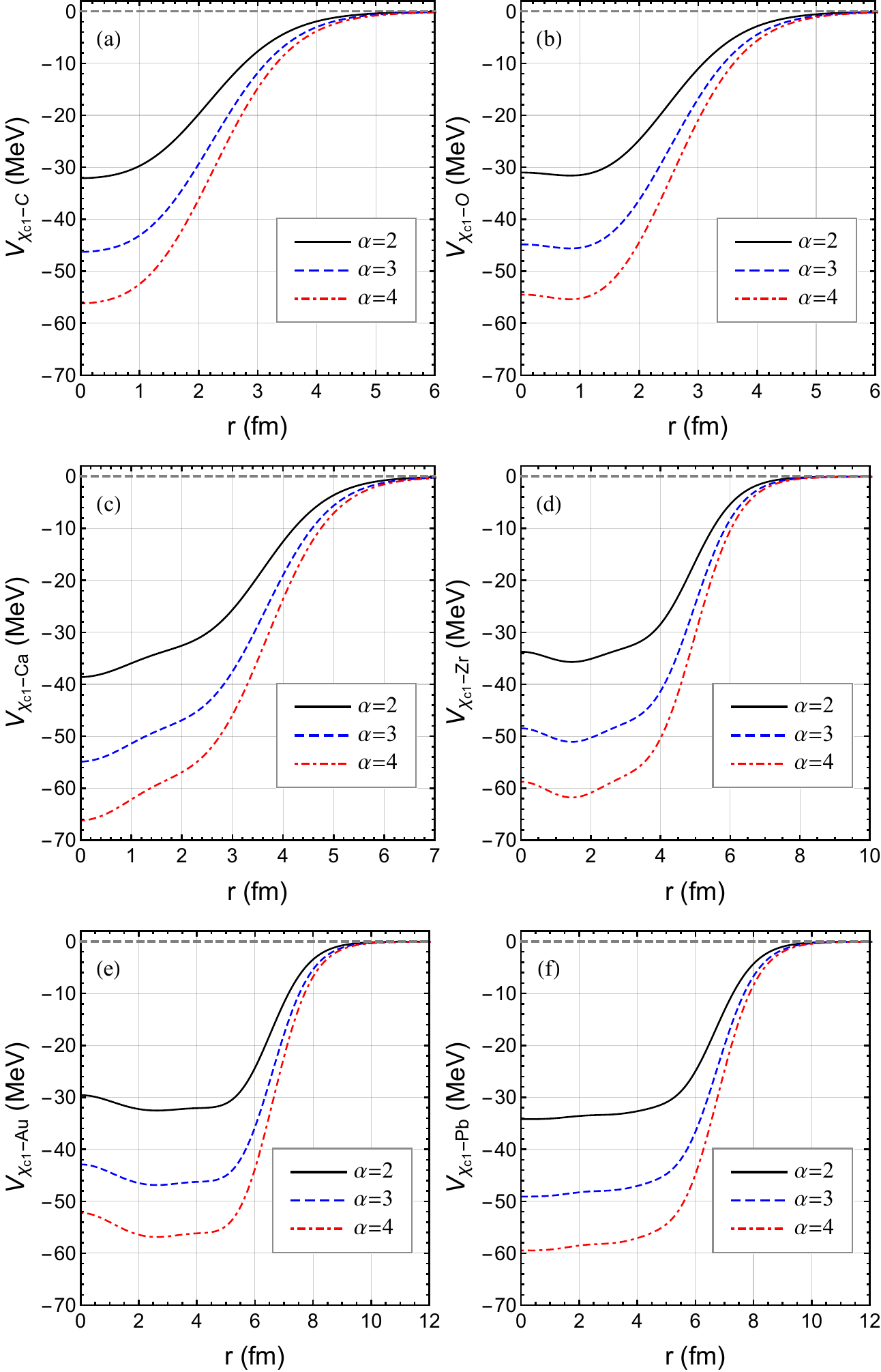}
    \caption{Radial distributions of the $\chi_{c1}$--nucleus potential
    $V_{\chi_{c1}-A}(r)$ for several nuclei, calculated with different
    values of the cutoff parameter $\alpha$. Panels (a)--(f) correspond
    to $^{12}$C, $^{16}$O, $^{40}$Ca, $^{90}$Zr, $^{197}$Au, and
    $^{208}$Pb, respectively.}
    \label{potential}
\end{figure}

Using the theoretical framework described above, we then calculate
the $\chi_{cJ}$--nucleus potentials $V_{\chi_{cJ}-A}(r)$.  The finite nuclei considered here are $^{12}\mathrm{C}$, $^{16}\mathrm{O}$, $^{40}\mathrm{Ca}$, $^{90}\mathrm{Zr}$, $^{197}\mathrm{Au}$, and $^{208}\mathrm{Pb}$. Since these potentials are obtained within the local density approximation, the radial dependence for different $\chi_{cJ}(1P)$ states is similar, while their depths differ according to the corresponding in-medium mass shifts. Therefore, in the following we take $\chi_{c1}(1P)$ as an example and show the calculated $V_{\chi_{c1}-A}(r)$ in Fig.~\ref{potential}.

Figure~\ref{potential} shows that increasing the cutoff parameter $\alpha$ mainly deepens the attractive potential, while the overall radial shape remains similar. With increasing nuclear mass number, the potential extends over a larger spatial region and develops a wider interior region.

Using the $\chi_{cJ}$--nucleus potentials constructed within the local density approximation, we calculate the bound state energies $E_{n\ell}^{(J,\,A)}$ by solving the Klein--Gordon equation, Eq.~(\ref{eq:kg}). The resulting spectra for $\chi_{c0}(1P)$, $\chi_{c1}(1P)$, and $\chi_{c2}(1P)$ are shown in Fig.~\ref{fig:chicJ-level-spectra}.

\tikzset{
    alpha2/.style={
        black!75!black,
        solid,
        line width=0.85pt,
        line cap=butt
    },
    alpha3/.style={
        blue!75!black,
        dash pattern=on 3pt off 1.6pt,
        line width=0.85pt,
        line cap=butt
    },
    alpha4/.style={
        red!70!black,
        dash pattern=on 0.6pt off 0.9pt on 2.4pt off 0.9pt,
        line width=0.85pt,
        line cap=round
    },
    toplabel/.style={
        font=\scriptsize\bfseries,
        anchor=base,
        inner sep=0pt,
        yshift=8pt,
        text height=1.45ex,
        text depth=0.25ex
    },
    bottomlabel/.style={
        font=\small,
        anchor=north,
        inner sep=0pt,
        yshift=-7pt
    },
    groupseparator/.style={
        gray!28,
        line width=0.35pt,
        densely dotted
    },
    paneltag/.style={
        font=\small\bfseries,
        anchor=south west,
        inner sep=1.5pt,
        fill=white,
        fill opacity=0.88,
        text opacity=1
    }
}

\newcommand{\Level}[3]{%
    \draw[#1]
    ([xshift=-6.8pt]axis cs:#2,#3) --
    ([xshift= 6.8pt]axis cs:#2,#3);
}
\newcommand{\ChicLevel}[3]{%
			\draw[#1]
			([xshift=-6.8pt]axis cs:#2,#3) --
			([xshift= 6.8pt]axis cs:#2,#3);
		}
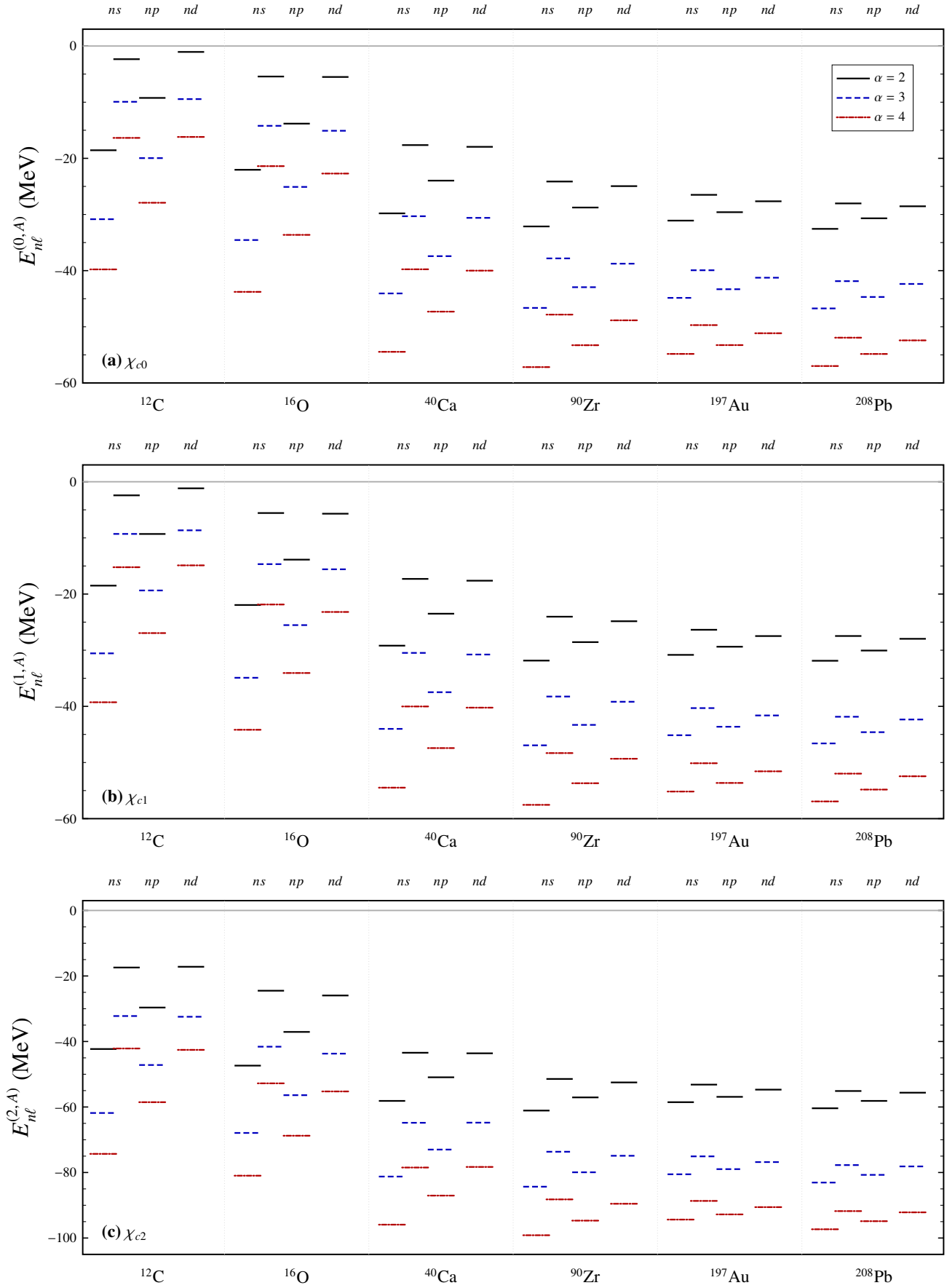
\begin{figure*}[p]
	\centering
	
	\begin{tikzpicture}[trim axis left, trim axis right]
		\begin{axis}[
			width=0.92\textwidth,
            height=0.25\textheight,
			scale only axis,
			axis lines=box,
			axis line style={
				black,
				line width=0.75pt
			},
			xmin=-1.35,
			xmax=15.35,
			ymin=-60,
			ymax=3,
			ylabel={$E_{n\ell}^{(0,\,A)}$ (MeV)},
			ylabel style={
				font=\large
			},
			xtick=\empty,
			ytick={-60,-40,-20,0},
			minor y tick num=3,
			tick align=inside,
			major tick length=0.10cm,
			minor tick length=0.06cm,
			every major tick/.append style={
				black,
				line width=0.60pt
			},
			every minor tick/.append style={
				black,
				line width=0.40pt
			},
			yticklabel style={
				font=\scriptsize
			},
			clip=false,
			legend style={
				draw=black,
				fill=white,
				line width=0.35pt,
				font=\scriptsize,
				at={(0.965,0.90)},
				anchor=north east,
				inner sep=2.0pt
			},
			legend cell align=left
			]
			
			\node[
			anchor=north west,
			font=\small\bfseries
			] at (rel axis cs:0.015,0.10) {(a)$\ \chi_{c0}$};
			
			\draw[
			gray!65,
			line width=0.70pt,
			]
			(axis cs:-1.35,0) -- (axis cs:15.35,0);
			
			\draw[groupseparator] (axis cs:1.40,-60) -- (axis cs:1.40,3);
			\draw[groupseparator] (axis cs:4.20,-60) -- (axis cs:4.20,3);
			\draw[groupseparator] (axis cs:7.00,-60) -- (axis cs:7.00,3);
			\draw[groupseparator] (axis cs:9.80,-60) -- (axis cs:9.80,3);
			\draw[groupseparator] (axis cs:12.60,-60) -- (axis cs:12.60,3);
			
			
			\Level{alpha2}{-0.95}{-18.56}
			\Level{alpha3}{-0.95}{-30.84}
			\Level{alpha4}{-0.95}{-39.78}
			
			\Level{alpha2}{-0.50}{-2.34}
			\Level{alpha3}{-0.50}{-9.93}
			\Level{alpha4}{-0.50}{-16.36}
			
			\Level{alpha2}{0.00}{-9.24}
			\Level{alpha3}{0.00}{-19.96}
			\Level{alpha4}{0.00}{-27.91}
			
			\Level{alpha2}{0.75}{-1.05}
			\Level{alpha3}{0.75}{-9.45}
			\Level{alpha4}{0.75}{-16.19}
			
			
			\Level{alpha2}{1.85}{-22.04}
			\Level{alpha3}{1.85}{-34.55}
			\Level{alpha4}{1.85}{-43.78}
			
			\Level{alpha2}{2.30}{-5.43}
			\Level{alpha3}{2.30}{-14.21}
			\Level{alpha4}{2.30}{-21.39}
			
			\Level{alpha2}{2.80}{-13.82}
			\Level{alpha3}{2.80}{-25.09}
			\Level{alpha4}{2.80}{-33.63}
			
			\Level{alpha2}{3.55}{-5.51}
			\Level{alpha3}{3.55}{-15.09}
			\Level{alpha4}{3.55}{-22.71}
			
			
			\Level{alpha2}{4.65}{-29.80}
			\Level{alpha3}{4.65}{-44.06}
			\Level{alpha4}{4.65}{-54.46}
			
			\Level{alpha2}{5.10}{-17.64}
			\Level{alpha3}{5.10}{-30.31}
			\Level{alpha4}{5.10}{-39.76}
			
			\Level{alpha2}{5.60}{-23.97}
			\Level{alpha3}{5.60}{-37.42}
			\Level{alpha4}{5.60}{-47.30}
			
			\Level{alpha2}{6.35}{-17.95}
			\Level{alpha3}{6.35}{-30.60}
			\Level{alpha4}{6.35}{-40.00}
			
			
			\Level{alpha2}{7.45}{-32.13}
			\Level{alpha3}{7.45}{-46.64}
			\Level{alpha4}{7.45}{-57.18}
			
			\Level{alpha2}{7.90}{-24.13}
			\Level{alpha3}{7.90}{-37.82}
			\Level{alpha4}{7.90}{-47.83}
			
			\Level{alpha2}{8.40}{-28.76}
			\Level{alpha3}{8.40}{-42.95}
			\Level{alpha4}{8.40}{-53.28}
			
			\Level{alpha2}{9.15}{-24.95}
			\Level{alpha3}{9.15}{-38.76}
			\Level{alpha4}{9.15}{-48.85}
			
			
			\Level{alpha2}{10.25}{-31.09}
			\Level{alpha3}{10.25}{-44.85}
			\Level{alpha4}{10.25}{-54.83}
			
			\Level{alpha2}{10.70}{-26.50}
			\Level{alpha3}{10.70}{-39.93}
			\Level{alpha4}{10.70}{-49.70}
			
			\Level{alpha2}{11.20}{-29.59}
			\Level{alpha3}{11.20}{-43.31}
			\Level{alpha4}{11.20}{-53.26}
			
			\Level{alpha2}{11.95}{-27.65}
			\Level{alpha3}{11.95}{-41.26}
			\Level{alpha4}{11.95}{-51.15}
			
			
			\Level{alpha2}{13.05}{-32.56}
			\Level{alpha3}{13.05}{-46.73}
			\Level{alpha4}{13.05}{-56.99}
			
			\Level{alpha2}{13.50}{-28.03}
			\Level{alpha3}{13.50}{-41.87}
			\Level{alpha4}{13.50}{-51.94}
			
			\Level{alpha2}{14.00}{-30.68}
			\Level{alpha3}{14.00}{-44.69}
			\Level{alpha4}{14.00}{-54.84}
			
			\Level{alpha2}{14.75}{-28.54}
			\Level{alpha3}{14.75}{-42.37}
			\Level{alpha4}{14.75}{-52.42}
			
			\node[toplabel] at (axis cs:-0.725,3) {$ns$};
			\node[toplabel] at (axis cs:0.000,3)  {$np$};
			\node[toplabel] at (axis cs:0.750,3)  {$nd$};
			
			\node[toplabel] at (axis cs:2.075,3) {$ns$};
			\node[toplabel] at (axis cs:2.800,3) {$np$};
			\node[toplabel] at (axis cs:3.550,3) {$nd$};
			
			\node[toplabel] at (axis cs:4.875,3) {$ns$};
			\node[toplabel] at (axis cs:5.600,3) {$np$};
			\node[toplabel] at (axis cs:6.350,3) {$nd$};
			
			\node[toplabel] at (axis cs:7.675,3) {$ns$};
			\node[toplabel] at (axis cs:8.400,3) {$np$};
			\node[toplabel] at (axis cs:9.150,3) {$nd$};
			
			\node[toplabel] at (axis cs:10.475,3) {$ns$};
			\node[toplabel] at (axis cs:11.200,3) {$np$};
			\node[toplabel] at (axis cs:11.950,3) {$nd$};
			
			\node[toplabel] at (axis cs:13.275,3) {$ns$};
			\node[toplabel] at (axis cs:14.000,3) {$np$};
			\node[toplabel] at (axis cs:14.750,3) {$nd$};
			
			\node[bottomlabel] at (axis cs:0.00,-60)  {$^{12}$C};
			\node[bottomlabel] at (axis cs:2.80,-60)  {$^{16}$O};
			\node[bottomlabel] at (axis cs:5.60,-60)  {$^{40}$Ca};
			\node[bottomlabel] at (axis cs:8.40,-60)  {$^{90}$Zr};
			\node[bottomlabel] at (axis cs:11.20,-60) {$^{197}$Au};
			\node[bottomlabel] at (axis cs:14.00,-60) {$^{208}$Pb};
			
			\addlegendimage{alpha2}
			\addlegendentry{$\alpha=2$}
			
			\addlegendimage{alpha3}
			\addlegendentry{$\alpha=3$}
			
			\addlegendimage{alpha4}
			\addlegendentry{$\alpha=4$}
			
		\end{axis}
	\end{tikzpicture}
	
	\par\vspace{4.5mm}

	\begin{tikzpicture}[trim axis left, trim axis right]
	\begin{axis}[
		width=0.92\textwidth,
        height=0.25\textheight,
		scale only axis,
		axis lines=box,
		axis line style={
			black,
			line width=0.75pt
		},
		xmin=-1.35,
		xmax=15.35,
		ymin=-60,
		ymax=3,
		ylabel={$E_{n\ell}^{(1,\,A)}$ (MeV)},
		ylabel style={
			font=\large
		},
		xtick=\empty,
		ytick={-60,-40,-20,0},
		minor y tick num=3,
		tick align=inside,
		major tick length=0.10cm,
		minor tick length=0.06cm,
		every major tick/.append style={
			black,
			line width=0.60pt
		},
		every minor tick/.append style={
			black,
			line width=0.40pt
		},
		yticklabel style={
			font=\scriptsize
		},
		clip=false,
		legend style={
			draw=black,
			fill=white,
			line width=0.35pt,
			font=\scriptsize,
			at={(0.965,0.90)},
			anchor=north east,
			inner sep=2.0pt
		},
		legend cell align=left
		]
		
		\node[
		anchor=north west,
		font=\small\bfseries
		] at (rel axis cs:0.015,0.10) {(b)$\ \chi_{c1}$};
		
		\draw[
		gray!65,
		line width=0.70pt,
		]
		(axis cs:-1.35,0) -- (axis cs:15.35,0);
		
		\draw[groupseparator] (axis cs:1.40,-60) -- (axis cs:1.40,3);
		\draw[groupseparator] (axis cs:4.20,-60) -- (axis cs:4.20,3);
		\draw[groupseparator] (axis cs:7.00,-60) -- (axis cs:7.00,3);
		\draw[groupseparator] (axis cs:9.80,-60) -- (axis cs:9.80,3);
		\draw[groupseparator] (axis cs:12.60,-60) -- (axis cs:12.60,3);
		
		\ChicLevel{alpha2}{-0.95}{-18.50}
		\ChicLevel{alpha3}{-0.95}{-30.56}
		\ChicLevel{alpha4}{-0.95}{-39.27}
		
		\ChicLevel{alpha2}{-0.50}{-2.42}
		\ChicLevel{alpha3}{-0.50}{-9.29}
		\ChicLevel{alpha4}{-0.50}{-15.22}
		
		\ChicLevel{alpha2}{0.00}{-9.30}
		\ChicLevel{alpha3}{0.00}{-19.36}
		\ChicLevel{alpha4}{0.00}{-26.95}
		
		\ChicLevel{alpha2}{0.75}{-1.17}
		\ChicLevel{alpha3}{0.75}{-8.65}
		\ChicLevel{alpha4}{0.75}{-14.89}
		
		\ChicLevel{alpha2}{1.85}{-21.95}
		\ChicLevel{alpha3}{1.85}{-34.91}
		\ChicLevel{alpha4}{1.85}{-44.17}
		
		\ChicLevel{alpha2}{2.30}{-5.57}
		\ChicLevel{alpha3}{2.30}{-14.67}
		\ChicLevel{alpha4}{2.30}{-21.86}
		
		\ChicLevel{alpha2}{2.80}{-13.87}
		\ChicLevel{alpha3}{2.80}{-25.53}
		\ChicLevel{alpha4}{2.80}{-34.06}
		
		\ChicLevel{alpha2}{3.55}{-5.69}
		\ChicLevel{alpha3}{3.55}{-15.59}
		\ChicLevel{alpha4}{3.55}{-23.19}
		
		\ChicLevel{alpha2}{4.65}{-29.19}
		\ChicLevel{alpha3}{4.65}{-44.01}
		\ChicLevel{alpha4}{4.65}{-54.48}
		
		\ChicLevel{alpha2}{5.10}{-17.30}
		\ChicLevel{alpha3}{5.10}{-30.48}
		\ChicLevel{alpha4}{5.10}{-40.02}
		
		\ChicLevel{alpha2}{5.60}{-23.50}
		\ChicLevel{alpha3}{5.60}{-37.48}
		\ChicLevel{alpha4}{5.60}{-47.43}
		
		\ChicLevel{alpha2}{6.35}{-17.62}
		\ChicLevel{alpha3}{6.35}{-30.77}
		\ChicLevel{alpha4}{6.35}{-40.23}
		
		\ChicLevel{alpha2}{7.45}{-31.85}
		\ChicLevel{alpha3}{7.45}{-46.94}
		\ChicLevel{alpha4}{7.45}{-57.54}
		
		\ChicLevel{alpha2}{7.90}{-24.03}
		\ChicLevel{alpha3}{7.90}{-38.25}
		\ChicLevel{alpha4}{7.90}{-48.32}
		
		\ChicLevel{alpha2}{8.40}{-28.56}
		\ChicLevel{alpha3}{8.40}{-43.30}
		\ChicLevel{alpha4}{8.40}{-53.69}
		
		\ChicLevel{alpha2}{9.15}{-24.84}
		\ChicLevel{alpha3}{9.15}{-39.18}
		\ChicLevel{alpha4}{9.15}{-49.33}
		
		\ChicLevel{alpha2}{10.25}{-30.83}
		\ChicLevel{alpha3}{10.25}{-45.14}
		\ChicLevel{alpha4}{10.25}{-55.17}
		
		\ChicLevel{alpha2}{10.70}{-26.36}
		\ChicLevel{alpha3}{10.70}{-40.31}
		\ChicLevel{alpha4}{10.70}{-50.13}
		
		\ChicLevel{alpha2}{11.20}{-29.38}
		\ChicLevel{alpha3}{11.20}{-43.63}
		\ChicLevel{alpha4}{11.20}{-53.64}
		
		\ChicLevel{alpha2}{11.95}{-27.48}
		\ChicLevel{alpha3}{11.95}{-41.62}
		\ChicLevel{alpha4}{11.95}{-51.57}
		
		\ChicLevel{alpha2}{13.05}{-31.88}
		\ChicLevel{alpha3}{13.05}{-46.60}
		\ChicLevel{alpha4}{13.05}{-56.93}
		
		\ChicLevel{alpha2}{13.50}{-27.47}
		\ChicLevel{alpha3}{13.50}{-41.85}
		\ChicLevel{alpha4}{13.50}{-51.98}
		
		\ChicLevel{alpha2}{14.00}{-30.06}
		\ChicLevel{alpha3}{14.00}{-44.60}
		\ChicLevel{alpha4}{14.00}{-54.82}
		
		\ChicLevel{alpha2}{14.75}{-27.97}
		\ChicLevel{alpha3}{14.75}{-42.34}
		\ChicLevel{alpha4}{14.75}{-52.45}
		
		\node[toplabel] at (axis cs:-0.725,3) {$ns$};
		\node[toplabel] at (axis cs:0.000,3)  {$np$};
		\node[toplabel] at (axis cs:0.750,3)  {$nd$};
		
		\node[toplabel] at (axis cs:2.075,3) {$ns$};
		\node[toplabel] at (axis cs:2.800,3) {$np$};
		\node[toplabel] at (axis cs:3.550,3) {$nd$};
		
		\node[toplabel] at (axis cs:4.875,3) {$ns$};
		\node[toplabel] at (axis cs:5.600,3) {$np$};
		\node[toplabel] at (axis cs:6.350,3) {$nd$};
		
		\node[toplabel] at (axis cs:7.675,3) {$ns$};
		\node[toplabel] at (axis cs:8.400,3) {$np$};
		\node[toplabel] at (axis cs:9.150,3) {$nd$};
		
		\node[toplabel] at (axis cs:10.475,3) {$ns$};
		\node[toplabel] at (axis cs:11.200,3) {$np$};
		\node[toplabel] at (axis cs:11.950,3) {$nd$};
		
		\node[toplabel] at (axis cs:13.275,3) {$ns$};
		\node[toplabel] at (axis cs:14.000,3) {$np$};
		\node[toplabel] at (axis cs:14.750,3) {$nd$};
		
		\node[bottomlabel] at (axis cs:0.00,-60)  {$^{12}$C};
		\node[bottomlabel] at (axis cs:2.80,-60)  {$^{16}$O};
		\node[bottomlabel] at (axis cs:5.60,-60)  {$^{40}$Ca};
		\node[bottomlabel] at (axis cs:8.40,-60)  {$^{90}$Zr};
		\node[bottomlabel] at (axis cs:11.20,-60) {$^{197}$Au};
		\node[bottomlabel] at (axis cs:14.00,-60) {$^{208}$Pb};
		
		
		
		
	\end{axis}
	\end{tikzpicture}
	
	\par\vspace{4.5mm}
	
	\begin{tikzpicture}[trim axis left, trim axis right]
		\begin{axis}[
			width=0.92\textwidth,
            height=0.25\textheight,
			scale only axis,
			axis lines=box,
			axis line style={
				black,
				line width=0.75pt
			},
			xmin=-1.35,
			xmax=15.35,
			ymin=-105,
			ymax=3,
			ylabel={$E_{n\ell}^{(2,\,A)}$ (MeV)},
			ylabel style={
				font=\large
			},
			xtick=\empty,
			ytick={-100,-80,-60,-40,-20,0},
			minor y tick num=3,
			tick align=inside,
			major tick length=0.10cm,
			minor tick length=0.06cm,
			every major tick/.append style={
				black,
				line width=0.60pt
			},
			every minor tick/.append style={
				black,
				line width=0.40pt
			},
			yticklabel style={
				font=\scriptsize
			},
			clip=false,
			legend style={
				draw=black,
				fill=white,
				line width=0.35pt,
				font=\scriptsize,
				at={(0.965,0.90)},
				anchor=north east,
				inner sep=2.0pt
			},
			legend cell align=left
			]
			
			\node[
			anchor=north west,
			font=\small\bfseries
			] at (rel axis cs:0.015,0.10) {(c)$\ \chi_{c2}$};
			
			\draw[
			gray!65,
			line width=0.70pt,
			]
			(axis cs:-1.35,0) -- (axis cs:15.35,0);
			
			\draw[groupseparator] (axis cs:1.40,-105) -- (axis cs:1.40,3);
			\draw[groupseparator] (axis cs:4.20,-105) -- (axis cs:4.20,3);
			\draw[groupseparator] (axis cs:7.00,-105) -- (axis cs:7.00,3);
			\draw[groupseparator] (axis cs:9.80,-105) -- (axis cs:9.80,3);
			\draw[groupseparator] (axis cs:12.60,-105) -- (axis cs:12.60,3);
			
			\ChicLevel{alpha2}{-0.95}{-42.30}
			\ChicLevel{alpha3}{-0.95}{-61.83}
			\ChicLevel{alpha4}{-0.95}{-74.31}
			
			\ChicLevel{alpha2}{-0.50}{-17.42}
			\ChicLevel{alpha3}{-0.50}{-32.23}
			\ChicLevel{alpha4}{-0.50}{-42.14}
			
			\ChicLevel{alpha2}{0.00}{-29.64}
			\ChicLevel{alpha3}{0.00}{-47.18}
			\ChicLevel{alpha4}{0.00}{-58.53}
			
			\ChicLevel{alpha2}{0.75}{-17.19}
			\ChicLevel{alpha3}{0.75}{-32.45}
			\ChicLevel{alpha4}{0.75}{-42.56}
			
			\ChicLevel{alpha2}{1.85}{-47.37}
			\ChicLevel{alpha3}{1.85}{-67.92}
			\ChicLevel{alpha4}{1.85}{-80.97}
			
			\ChicLevel{alpha2}{2.30}{-24.50}
			\ChicLevel{alpha3}{2.30}{-41.59}
			\ChicLevel{alpha4}{2.30}{-52.78}
			
			\ChicLevel{alpha2}{2.80}{-37.08}
			\ChicLevel{alpha3}{2.80}{-56.39}
			\ChicLevel{alpha4}{2.80}{-68.77}
			
			\ChicLevel{alpha2}{3.55}{-25.96}
			\ChicLevel{alpha3}{3.55}{-43.71}
			\ChicLevel{alpha4}{3.55}{-55.25}
			
			\ChicLevel{alpha2}{4.65}{-58.14}
			\ChicLevel{alpha3}{4.65}{-81.24}
			\ChicLevel{alpha4}{4.65}{-95.92}
			
			\ChicLevel{alpha2}{5.10}{-43.44}
			\ChicLevel{alpha3}{5.10}{-64.82}
			\ChicLevel{alpha4}{5.10}{-78.48}
			
			\ChicLevel{alpha2}{5.60}{-50.93}
			\ChicLevel{alpha3}{5.60}{-73.00}
			\ChicLevel{alpha4}{5.60}{-87.05}
			
			\ChicLevel{alpha2}{6.35}{-43.61}
			\ChicLevel{alpha3}{6.35}{-64.78}
			\ChicLevel{alpha4}{6.35}{-78.30}
			
			\ChicLevel{alpha2}{7.45}{-61.09}
			\ChicLevel{alpha3}{7.45}{-84.33}
			\ChicLevel{alpha4}{7.45}{-99.15}
			
			\ChicLevel{alpha2}{7.90}{-51.44}
			\ChicLevel{alpha3}{7.90}{-73.66}
			\ChicLevel{alpha4}{7.90}{-88.21}
			
			\ChicLevel{alpha2}{8.40}{-57.07}
			\ChicLevel{alpha3}{8.40}{-79.93}
			\ChicLevel{alpha4}{8.40}{-94.69}
			
			\ChicLevel{alpha2}{9.15}{-52.50}
			\ChicLevel{alpha3}{9.15}{-74.89}
			\ChicLevel{alpha4}{9.15}{-89.54}
			
			\ChicLevel{alpha2}{10.25}{-58.53}
			\ChicLevel{alpha3}{10.25}{-80.54}
			\ChicLevel{alpha4}{10.25}{-94.37}
			
			\ChicLevel{alpha2}{10.70}{-53.17}
			\ChicLevel{alpha3}{10.70}{-75.07}
			\ChicLevel{alpha4}{10.70}{-88.67}
			
			\ChicLevel{alpha2}{11.20}{-56.90}
			\ChicLevel{alpha3}{11.20}{-78.97}
			\ChicLevel{alpha4}{11.20}{-92.79}
			
			\ChicLevel{alpha2}{11.95}{-54.70}
			\ChicLevel{alpha3}{11.95}{-76.81}
			\ChicLevel{alpha4}{11.95}{-90.57}
			
			\ChicLevel{alpha2}{13.05}{-60.39}
			\ChicLevel{alpha3}{13.05}{-83.07}
			\ChicLevel{alpha4}{13.05}{-97.33}
			
			\ChicLevel{alpha2}{13.50}{-55.12}
			\ChicLevel{alpha3}{13.50}{-77.73}
			\ChicLevel{alpha4}{13.50}{-91.78}
			
			\ChicLevel{alpha2}{14.00}{-58.14}
			\ChicLevel{alpha3}{14.00}{-80.72}
			\ChicLevel{alpha4}{14.00}{-94.86}
			
			\ChicLevel{alpha2}{14.75}{-55.63}
			\ChicLevel{alpha3}{14.75}{-78.14}
			\ChicLevel{alpha4}{14.75}{-92.16}
			
			\node[toplabel] at (axis cs:-0.725,3) {$ns$};
			\node[toplabel] at (axis cs:0.000,3)  {$np$};
			\node[toplabel] at (axis cs:0.750,3)  {$nd$};
			
			\node[toplabel] at (axis cs:2.075,3) {$ns$};
			\node[toplabel] at (axis cs:2.800,3) {$np$};
			\node[toplabel] at (axis cs:3.550,3) {$nd$};
			
			\node[toplabel] at (axis cs:4.875,3) {$ns$};
			\node[toplabel] at (axis cs:5.600,3) {$np$};
			\node[toplabel] at (axis cs:6.350,3) {$nd$};
			
			\node[toplabel] at (axis cs:7.675,3) {$ns$};
			\node[toplabel] at (axis cs:8.400,3) {$np$};
			\node[toplabel] at (axis cs:9.150,3) {$nd$};
			
			\node[toplabel] at (axis cs:10.475,3) {$ns$};
			\node[toplabel] at (axis cs:11.200,3) {$np$};
			\node[toplabel] at (axis cs:11.950,3) {$nd$};
			
			\node[toplabel] at (axis cs:13.275,3) {$ns$};
			\node[toplabel] at (axis cs:14.000,3) {$np$};
			\node[toplabel] at (axis cs:14.750,3) {$nd$};
			
			\node[bottomlabel] at (axis cs:0.00,-105)  {$^{12}$C};
			\node[bottomlabel] at (axis cs:2.80,-105)  {$^{16}$O};
			\node[bottomlabel] at (axis cs:5.60,-105)  {$^{40}$Ca};
			\node[bottomlabel] at (axis cs:8.40,-105)  {$^{90}$Zr};
			\node[bottomlabel] at (axis cs:11.20,-105) {$^{197}$Au};
			\node[bottomlabel] at (axis cs:14.00,-105) {$^{208}$Pb};
			
			
			
			
		\end{axis}
	\end{tikzpicture}
	
	\caption{
		Bound state energies $E_{n\ell}^{(J,\,A)}$ of the
		$\chi_{cJ}$--nucleus systems for $\alpha=2$, 3, and 4.
		Panels (a), (b), and (c) correspond to $\chi_{c0}(1P)$,
		$\chi_{c1}(1P)$, and $\chi_{c2}(1P)$, respectively. The six groups in each panel correspond to $^{12}$C, $^{16}$O, $^{40}$Ca, $^{90}$Zr, $^{197}$Au, and $^{208}$Pb.
	}
	\label{fig:chicJ-level-spectra}
\end{figure*}

The corresponding numerical values of the bound state energies are listed in Table~\ref{tab:chic-binding} in Appendix~\ref{A}. As can be seen, all the considered $\chi_{cJ}$--nucleus systems have negative bound state energies, indicating that they can form nuclear bound states in all selected nuclei. This result is consistent with the result reported in Ref.~\cite{Mondal:2024kmo}. Even in the lightest system, bound states already appear. For example, the $\chi_{c0}$--$^{12}$C $1d$ state has a bound state energy of $-1.05$ MeV for $\alpha=2$. For medium and heavy nuclei, the bound state energies become substantially more negative. In the $\chi_{c0}$--$^{208}$Pb system, the $1s$ bound state energy changes from $-32.56$ MeV to $-56.99$ MeV when $\alpha$ increases from 2 to 4, while in the $\chi_{c2}$--$^{208}$Pb system it reaches $-97.33$ MeV for $\alpha=4$. These results suggest the possible formation of deeply bound $\chi_{cJ}$--nuclear states.

For a given nucleus, orbital state, and cutoff parameter $\alpha$, the bound state energies of the $\chi_{c0}(1P)$ and $\chi_{c1}(1P)$ states are almost identical. This near degeneracy originates from two related facts. First, the vacuum masses of $\chi_{c0}(1P)$ and $\chi_{c1}(1P)$ are close. Second, their in-medium mass shifts in nuclear matter are also nearly the same, as shown in Table~\ref{tab:mass-shift-chicj-rho0}. For instance, in the $^{12}$C system with $\alpha=2$, the $1s$ bound state energies of $\chi_{c0}(1P)$ and $\chi_{c1}(1P)$ are $-18.56$ MeV and $-18.50$ MeV, respectively. In the $^{208}$Pb system with $\alpha=4$, the corresponding values are $-56.99$ MeV and $-56.93$ MeV. By contrast, the $\chi_{c2}(1P)$ bound state energies are systematically more negative, reflecting the larger downward mass shift of $\chi_{c2}(1P)$ in nuclear matter.

The dependence on the nuclear mass number also shows a clear systematic trend. For a fixed $\chi_{cJ}(1P)$ state and cutoff parameter, most bound state energies become more negative as the nucleus changes from light to medium and heavy systems. For example, for the $\chi_{c0}(1P)$ $1s$ state with $\alpha=2$, the bound state energy changes from $-18.56$ MeV in $^{12}$C to $-29.80$ MeV in $^{40}$Ca and $-32.56$ MeV in $^{208}$Pb. This behavior reflects the increasing spatial extension and effective attraction of heavier nuclei. In the heavy-mass region, however, the lowest $1s$ bound state energies tend to become nearly saturated. For $\chi_{c0}(1P)$ with $\alpha=2$, the $1s$ bound state energies in $^{90}$Zr, $^{197}$Au, and $^{208}$Pb are $-32.13$, $-31.09$, and $-32.56$ MeV, respectively. For $\alpha=4$, the corresponding values are $-57.18$, $-54.83$, and $-56.99$ MeV. This approximate equality arises because the central density of medium and heavy nuclei is already close to normal nuclear matter density, so the bottom of the effective potential becomes nearly saturated. Since the $1s$ wave function is mainly concentrated in the nuclear interior, the $1s$ bound state energies of the same $\chi_{cJ}(1P)$ state become nearly degenerate among heavy nuclei.

We next turn to the relative level structure. For this purpose, we define the energy spacing measured from the lowest $1s$ level as
\begin{equation}
\Delta E_{n\ell}^{(J,A)}
=
E_{n\ell}^{(J,A)}-E_{1s}^{(J,A)} ,
\end{equation}
where $E_{n\ell}^{(J,A)}$ denotes the bound state energy defined in Eq.~(39). Compared with the absolute bound state energies, these energy spacings are much more stable against the variation of the cutoff parameter $\alpha$. For example, in the $\chi_{c0}$--$^{208}$Pb system, the $1s$ bound state energy changes from $-32.56$ MeV to $-56.99$ MeV when $\alpha$ is increased from 2 to 4. In contrast, the $1p$--$1s$ energy spacing changes only from about $1.88$ MeV to about $2.15$ MeV. Thus, the cutoff parameter mainly shifts the overall depth of the spectrum, while the relative spacings are only moderately affected.

A characteristic pattern appears among the low-lying states up to the $2s$ level. For a fixed nucleus, $\chi_{cJ}(1P)$ state, and cutoff parameter, the energy spacings of the $1p$, $1d$, and $2s$ levels from the $1s$ level approximately follow the ratio
\[
\Delta E_{1p}:\Delta E_{1d}:\Delta E_{2s}
\simeq
1:2:2 .
\]
This pattern is reminiscent of the three-dimensional harmonic-oscillator shell counting, where the excitation quantum number is
\begin{equation}
N_{n\ell}=2(n-1)+\ell .
\end{equation}
Thus the $1p$, $1d$, and $2s$ states correspond to $N_{n\ell}=1,2,2$, respectively. As a typical example, for $\chi_{c0}$--$^{40}$Ca with $\alpha=2$, the energy spacings obtained from Table~\ref{tab:chic-binding} are $\Delta E_{1p}=5.83~{\rm MeV}$, $\Delta E_{1d}=11.85~{\rm MeV}$, and $\Delta E_{2s}=12.16~{\rm MeV}$, which are close to the ratio $1:2:2$. This approximate oscillator-like pattern may be traced back to the local shape of the $\chi_{cJ}$--nucleus potential near the nuclear center. For a smooth attractive central potential, the leading radial correction around its minimum is approximately quadratic, giving rise to an oscillator-like ordering for the lowest states. This argument is only qualitative, however, because the realistic $\chi_{cJ}$--nucleus potential has a finite depth, a diffuse surface, and non-quadratic behavior away from the center. Consequently, the $1d$ and $2s$ levels are not exactly degenerate, and the oscillator-like regularity may gradually break down for higher excited states whose wave functions probe the surface and tail regions of the potential.

\begin{figure}[H]
\centering
\resizebox{0.75\columnwidth}{!}{%
\begin{tikzpicture}[
    >=Stealth,
    font=\small,
    axis/.style={->,line width=0.9pt},
    well/.style={line width=1.0pt}
]

\def\RA{5.2}
\def\Vbottom{-2.0}

\shade[
    top color=blue!12,
    bottom color=blue!45,
    draw=none
] (0,0) rectangle (\RA,\Vbottom);

\draw[well,blue!70!black] (0,0) -- (\RA,0);
\draw[well,blue!70!black] (\RA,0) -- (\RA,\Vbottom);
\draw[well,blue!70!black] (0,\Vbottom) -- (\RA,\Vbottom);

\draw[axis,black!70!black] (0,\Vbottom-0.45) -- (0,0.55)
    node[above] {$V_{\chi_{cJ}-A}(r)$};
\draw[axis,black!70!black] (0,0) -- (\RA+0.75,0)
    node[right] {$r$};

\node[left] at (0,0) {$0$};
\node[left] at (0,\Vbottom) {$-V_0$};
\node[above] at (\RA,0.05) {$R_A$};

\node[align=center,text=black] at ({0.5*\RA},{0.5*\Vbottom})
    {finite attractive\\ potential well};

\end{tikzpicture}%
}
\caption{
Schematic illustration of the finite-well approximation to the
$\chi_{cJ}$--nucleus attractive potential.
}
\label{finite-well-picture}
\end{figure}
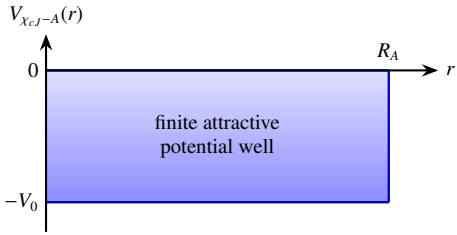

The dependence of the energy spacings on the nuclear mass number A has a different origin. It is mainly controlled by the finite spatial extension of the nuclear potential. To see this, one may approximate the attractive potential by a finite spherical well, as shown in Fig.~\ref{finite-well-picture},
\begin{equation}
V_{\chi_{cJ}-A}(r)
=
\begin{cases}
-V_0, & r<R_A,\\
0, & r>R_A,
\end{cases}
\end{equation}
where $R_A$ is the nuclear radius. Inside the well, the Klein--Gordon
equation becomes
\begin{equation}
	\left[
	-\nabla^2+\left(\mu_A^{(J)}\right)^2-2\mu_A^{(J)} V_0
	\right]\phi(\bm r)
	=
	\mathcal E_{n\ell}^2\phi(\bm r),
\end{equation}
where $\mu_A^{(J)}$ is the reduced mass of the $\chi_{cJ}$--nucleus system. For the purpose of estimating the level spacing, we further use the hard-wall approximation for a deep well, so that the radial wave function satisfies approximately $j_\ell(kR_A)=0$. Thus one has
\begin{equation}
	k_{n\ell}\simeq \frac{x_{n\ell}}{R_A},
\end{equation}
where $x_{n\ell}$ denotes the corresponding zero of the spherical Bessel function $j_\ell(x)$. Therefore, the bound state energy is
\begin{equation}
	E_{n\ell}^{(J,\,A)}
	\simeq
	\sqrt{
	\left(\mu_A^{(J)}\right)^2
	-2\mu_A^{(J)} V_0
	+\frac{x_{n\ell}^2}{R_A^2}
	}
	-\mu_A^{(J)} .
\end{equation} 
Taking the difference with respect to the $1s$ level, one obtains
\begin{equation}
\Delta E_{n\ell}^{(J,A)}
\simeq
\frac{
x_{n\ell}^2-x_{1s}^2
}{
2\mu_A^{(J)}R_A^2
}.
\end{equation}
Since the nuclear radius approximately follows
\begin{equation}
R_A= r_0 A^{1/3},
\end{equation}
and the reduced mass $\mu_A^{(J)}$ varies only weakly for medium and heavy nuclei, the leading nuclear-mass dependence of the energy spacing is
\begin{equation}
\Delta E_{n\ell}^{(J,A)}
\propto
A^{-2/3}.
\end{equation}
This finite-radius quantization picture explains why heavier nuclei have denser spectra.

\pgfplotsset{
	threepanel/.style={
		width=4.85cm,
		height=4.15cm,
		scale only axis,		
		xmin=0,
		xmax=0.4,
		ymin=0,
		ymax=35,		
		xtick={0,0.1,0.2,0.3,0.4},
		ytick={0,5,10,15,20,25,30,35},		
		minor x tick num=4,
		minor y tick num=4,		
		xtick pos=both,
		ytick pos=both,
		tick align=inside,		
		axis line style={
			black,
			line width=0.8pt
		},		
		major tick style={
			black,
			line width=0.65pt
		},		
		minor tick style={
			black,
			line width=0.4pt
		},		
		major tick length=3.5pt,
		minor tick length=1.8pt,		
		xlabel={$N_{n\ell}A^{-2/3}$},		
		label style={
			font=\small
		},		
		tick label style={
			font=\scriptsize
		},		
		xticklabel style={
			/pgf/number format/fixed,
			/pgf/number format/fixed zerofill,
			/pgf/number format/precision=1
		},		
		yticklabel style={
			/pgf/number format/fixed,
			/pgf/number format/precision=0
		},		
		clip=true
	},	
	lineA/.style={
		black,
		line width=0.95pt
	},	
	lineB/.style={
		blue,
		line width=0.95pt,
		dash pattern=on 4pt off 2.5pt
	},	
	lineC/.style={
		red,
		line width=0.95pt,
		dash pattern=on 1pt off 1.1pt on 3pt off 1.1pt
	},	
	markA/.style={
		only marks,
		black,
		mark=*,
		mark size=1.55pt,
		mark options={
			solid,
			fill=black,
			draw=black
		}
	},	
	markB/.style={
		only marks,
		blue,
		mark=square*,
		mark size=1.55pt,
		mark options={
			solid,
			fill=blue,
			draw=blue
		}
	},	
	markC/.style={
		only marks,
		red,
		mark=diamond*,
		mark size=1.75pt,
		mark options={
			solid,
			fill=red,
			draw=red
		}
	}
}

\begin{figure*}[!t]
	\centering
	\begin{tikzpicture}
		
		\begin{axis}[
			threepanel,
			at={(0cm,0cm)},
			anchor=south west,			
			ylabel={$\Delta E_{n\ell}\;(\mathrm{MeV})$},			
			legend style={
				at={(0.95,0.05)},
				anchor=south east,
				draw=gray!60,
				fill=white,
				line width=0.35pt,
				font=\scriptsize,
				inner xsep=3pt,
				inner ysep=2pt,
				row sep=-1pt
			},			
			legend cell align=left
			]
			
			\addplot[
			lineA,
			domain=0:0.4,
			samples=2,
			forget plot
			]
			{51.5602*x};
			
			\addplot[
			lineB,
			domain=0:0.4,
			samples=2,
			forget plot
			]
			{61.8292*x};
			
			\addplot[
			lineC,
			domain=0:0.4,
			samples=2,
			forget plot
			]
			{67.6054*x};
			
			\addplot[
			markA,
			forget plot
			]
			coordinates {
				(0.1907857071,  9.32)
				(0.1574901312,  8.22)
				(0.0854987973,  5.83)
				(0.0497933861,  3.37)
				(0.0295362836,  1.50)
				(0.0284855391,  1.88)
				
				(0.3815714142, 17.51)
				(0.3149802625, 16.53)
				(0.1709975947, 11.85)
				(0.0995867721,  7.18)
				(0.0590725672,  3.44)
				(0.0569710782,  4.02)
				
				(0.3815714142, 16.22)
				(0.3149802625, 16.61)
				(0.1709975947, 12.16)
				(0.0995867721,  8.00)
				(0.0590725672,  4.59)
				(0.0569710782,  4.53)
			};
			
			\addplot[
			markB,
			forget plot
			]
			coordinates {
				(0.1907857071, 10.88)
				(0.1574901312,  9.46)
				(0.0854987973,  6.64)
				(0.0497933861,  3.69)
				(0.0295362836,  1.54)
				(0.0284855391,  2.04)
				
				(0.3815714142, 21.39)
				(0.3149802625, 19.46)
				(0.1709975947, 13.46)
				(0.0995867721,  7.88)
				(0.0590725672,  3.59)
				(0.0569710782,  4.36)
				
				(0.3815714142, 20.91)
				(0.3149802625, 20.34)
				(0.1709975947, 13.75)
				(0.0995867721,  8.82)
				(0.0590725672,  4.92)
				(0.0569710782,  4.86)
			};
			
			\addplot[
			markC,
			forget plot
			]
			coordinates {
				(0.1907857071, 11.87)
				(0.1574901312, 10.15)
				(0.0854987973,  7.16)
				(0.0497933861,  3.90)
				(0.0295362836,  1.57)
				(0.0284855391,  2.15)
				
				(0.3815714142, 23.59)
				(0.3149802625, 21.07)
				(0.1709975947, 14.46)
				(0.0995867721,  8.33)
				(0.0590725672,  3.68)
				(0.0569710782,  4.57)
				
				(0.3815714142, 23.42)
				(0.3149802625, 22.39)
				(0.1709975947, 14.70)
				(0.0995867721,  9.35)
				(0.0590725672,  5.13)
				(0.0569710782,  5.05)
			};
			
			\addlegendimage{
				lineA,
				mark=*,
				mark size=1.55pt,
				mark options={
					solid,
					fill=black,
					draw=black
				}
			}
			\addlegendentry{$\alpha=2$}
			
			\addlegendimage{
				lineB,
				mark=square*,
				mark size=1.55pt,
				mark options={
					solid,
					fill=blue,
					draw=blue
				}
			}
			\addlegendentry{$\alpha=3$}
			
			\addlegendimage{
				lineC,
				mark=diamond*,
				mark size=1.75pt,
				mark options={
					solid,
					fill=red,
					draw=red
				}
			}
			\addlegendentry{$\alpha=4$}
			
			\node[
			anchor=north west,
			font=\bfseries\normalsize
			]
			at (rel axis cs:0.035,0.965) {(a) $\chi_{c0}$};
			
		\end{axis}

		\begin{axis}[
			threepanel,			
			at={(5.30cm,0cm)},
			anchor=south west,			
			yticklabels=\empty
			]
			
			\addplot[
			lineA,
			domain=0:0.4,
			samples=2,
			forget plot
			]
			{50.7987*x};
			
			\addplot[
			lineB,
			domain=0:0.4,
			samples=2,
			forget plot
			]
			{62.094*x};
			
			\addplot[
			lineC,
			domain=0:0.4,
			samples=2,
			forget plot
			]
			{68.2642*x};
			
			\addplot[
			markA,
			forget plot
			]
			coordinates {
				(0.1907857071,  9.20)
				(0.1574901312,  8.08)
				(0.0854987973,  5.69)
				(0.0497933861,  3.29)
				(0.0295362836,  1.45)
				(0.0284855391,  1.82)
				
				(0.3815714142, 17.33)
				(0.3149802625, 16.26)
				(0.1709975947, 11.57)
				(0.0995867721,  7.01)
				(0.0590725672,  3.35)
				(0.0569710782,  3.91)
				
				(0.3815714142, 16.08)
				(0.3149802625, 16.38)
				(0.1709975947, 11.89)
				(0.0995867721,  7.82)
				(0.0590725672,  4.47)
				(0.0569710782,  4.41)
			};
			
			\addplot[
			markB,
			forget plot
			]
			coordinates {
				(0.1907857071, 11.20)
				(0.1574901312,  9.38)
				(0.0854987973,  6.53)
				(0.0497933861,  3.64)
				(0.0295362836,  1.51)
				(0.0284855391,  2.00)
				
				(0.3815714142, 21.91)
				(0.3149802625, 19.32)
				(0.1709975947, 13.24)
				(0.0995867721,  7.76)
				(0.0590725672,  3.52)
				(0.0569710782,  4.26)
				
				(0.3815714142, 21.27)
				(0.3149802625, 20.24)
				(0.1709975947, 13.53)
				(0.0995867721,  8.69)
				(0.0590725672,  4.83)
				(0.0569710782,  4.75)
			};
			
			\addplot[
			markC,
			forget plot
			]
			coordinates {
				(0.1907857071, 12.32)
				(0.1574901312, 10.11)
				(0.0854987973,  7.05)
				(0.0497933861,  3.85)
				(0.0295362836,  1.53)
				(0.0284855391,  2.11)
				
				(0.3815714142, 24.38)
				(0.3149802625, 20.98)
				(0.1709975947, 14.25)
				(0.0995867721,  8.21)
				(0.0590725672,  3.60)
				(0.0569710782,  4.48)
				
				(0.3815714142, 24.05)
				(0.3149802625, 22.31)
				(0.1709975947, 14.46)
				(0.0995867721,  9.22)
				(0.0590725672,  5.04)
				(0.0569710782,  4.95)
			};
			
			\node[
			anchor=north west,
			font=\bfseries\normalsize
			]
			at (rel axis cs:0.035,0.965) {(b) $\chi_{c1}$};
			
		\end{axis}

		\begin{axis}[
			threepanel,			
			at={(10.60cm,0cm)},
			anchor=south west,			
			yticklabels=\empty
			]
			
			\addplot[
			lineA,
			domain=0:0.4,
			samples=2,
			forget plot
			]
			{70.1992*x};
			
			\addplot[
			lineB,
			domain=0:0.4,
			samples=2,
			forget plot
			]
			{80.7632*x};
			
			\addplot[
			lineC,
			domain=0:0.4,
			samples=2,
			forget plot
			]
			{86.5547*x};
			
			\addplot[
			markA,
			forget plot
			]
			coordinates {
				(0.1907857071, 12.66)
				(0.1574901312, 10.29)
				(0.0854987973,  7.21)
				(0.0497933861,  4.02)
				(0.0295362836,  1.63)
				(0.0284855391,  2.25)
				
				(0.3815714142, 25.11)
				(0.3149802625, 21.41)
				(0.1709975947, 14.53)
				(0.0995867721,  8.59)
				(0.0590725672,  3.83)
				(0.0569710782,  4.76)
				
				(0.3815714142, 24.88)
				(0.3149802625, 22.87)
				(0.1709975947, 14.70)
				(0.0995867721,  9.65)
				(0.0590725672,  5.36)
				(0.0569710782,  5.27)
			};
			
			\addplot[
			markB,
			forget plot
			]
			coordinates {
				(0.1907857071, 14.65)
				(0.1574901312, 11.53)
				(0.0854987973,  8.24)
				(0.0497933861,  4.40)
				(0.0295362836,  1.57)
				(0.0284855391,  2.35)
				
				(0.3815714142, 29.38)
				(0.3149802625, 24.21)
				(0.1709975947, 16.46)
				(0.0995867721,  9.44)
				(0.0590725672,  3.73)
				(0.0569710782,  4.93)
				
				(0.3815714142, 29.60)
				(0.3149802625, 26.33)
				(0.1709975947, 16.42)
				(0.0995867721, 10.67)
				(0.0590725672,  5.47)
				(0.0569710782,  5.34)
			};
			
			\addplot[
			markC,
			forget plot
			]
			coordinates {
				(0.1907857071, 15.78)
				(0.1574901312, 12.20)
				(0.0854987973,  8.87)
				(0.0497933861,  4.46)
				(0.0295362836,  1.58)
				(0.0284855391,  2.47)
				
				(0.3815714142, 31.75)
				(0.3149802625, 25.72)
				(0.1709975947, 17.62)
				(0.0995867721,  9.61)
				(0.0590725672,  3.80)
				(0.0569710782,  5.17)
				
				(0.3815714142, 32.17)
				(0.3149802625, 28.19)
				(0.1709975947, 17.44)
				(0.0995867721, 10.94)
				(0.0590725672,  5.70)
				(0.0569710782,  5.55)
			};
			
			\node[
			anchor=north west,
			font=\bfseries\normalsize
			]
			at (rel axis cs:0.035,0.965) {(c) $\chi_{c2}$};
			
		\end{axis}
		
	\end{tikzpicture}
	
	\caption{Scaling behavior of the energy spacings $\Delta E_{n\ell}^{(J,A)}=E_{n\ell}^{(J,A)}-E_{1s}^{(J,A)}$ as functions of $N_{n\ell}A^{-2/3}$, where $N_{n\ell}=2(n-1)+\ell$. Panels (a), (b), and (c) correspond to $\chi_{c0}$, $\chi_{c1}$, and $\chi_{c2}$ nuclear bound states, respectively. Black circles with solid lines, blue squares with dashed lines, and red diamonds with dot-dashed lines denote the results for $\alpha=2$, 3, and 4, respectively. The symbols denote the numerical $1p-1s$, $1d-1s$, and $2s-1s$ energy spacings for the six nuclei, while the lines are origin-constrained linear fits based on Eq.~(49). The uncentered coefficients of determination of all the fits are larger than 0.95.}
	\label{line}
\end{figure*}
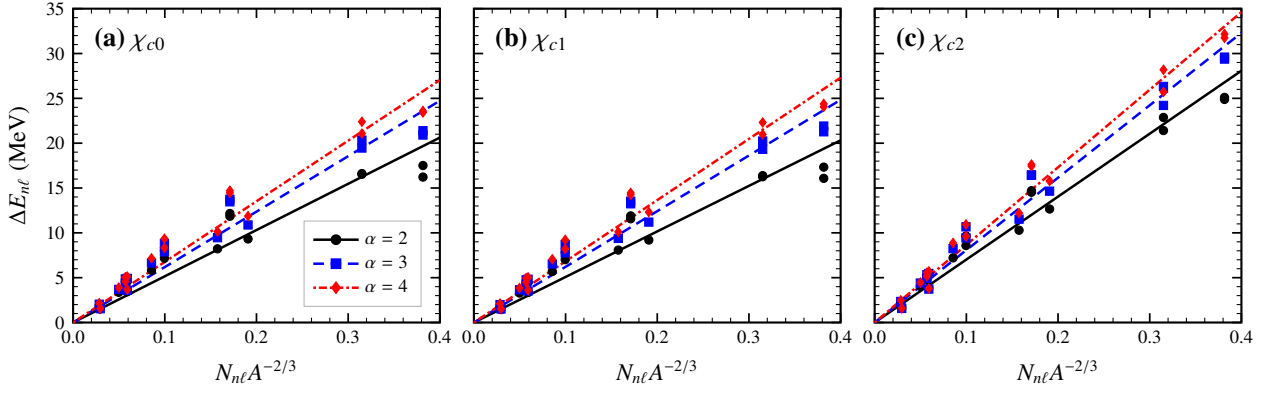

Combining the oscillator-like counting of the low-lying states with the finite-radius scaling, the numerical results can be summarized by the approximate relation
\begin{equation}
\Delta E_{n\ell}^{(J,A)}
\propto
N_{n\ell}A^{-2/3},
\end{equation}
where $N_{n\ell}=2(n-1)+\ell$. As shown in Fig.~\ref{line}, panels (a), (b), and (c) correspond to the $\chi_{c0}(1P)$, $\chi_{c1}(1P)$, and $\chi_{c2}(1P)$ nuclear bound states, respectively. The black circles with solid lines, blue squares with dashed lines, and red diamonds with dot-dashed lines denote the results for $\alpha=2$, 3, and 4, respectively. For each fixed $\chi_{cJ}(1P)$ state and cutoff parameter $\alpha$, the points associated with the $1p-1s$, $1d-1s$, and $2s-1s$ energy spacings are approximately distributed along a straight line when plotted as functions of $N_{n\ell}A^{-2/3}$. The quality of the origin-constrained linear fits is quantified by
the uncentered coefficient of determination,
\begin{equation}
R_{\rm unc}^{2}=1-\frac{\sum_i\left(\Delta E_i-\Delta E_i^{\rm fit}\right)^2}{\sum_i\left(\Delta E_i\right)^2},
\end{equation}
where $\Delta E_i^{\rm fit}$ denotes the value obtained from the corresponding origin-constrained linear fit. The resulting $R_{\rm unc}^{2}$ values are all larger than 0.95, indicating that the energy spacings follow a robust nearly linear scaling pattern. Deviations from exact linearity are expected because the actual $\chi_{cJ}$--nucleus potential is neither an ideal harmonic oscillator near the center nor an infinitely sharp square well at the surface. Nevertheless, the overall behavior shows that the relative energy spacings are mainly governed by the finite nuclear radius and can be organized by the simple scaling variable $N_{n\ell}A^{-2/3}$.

Finally, we comment on the possible widths and experimental signatures of the predicted \(\chi_{cJ}\)--nucleus bound states.
In contrast to the $J/\psi$--nucleus case, where the natural width of the free $J/\psi$ is tiny and the open-charm dissociation channel $J/\psi+N\to \Lambda_c^+ + D^-$ is kinematically forbidden for a nearly recoilless $J/\psi$, the corresponding process $\chi_{cJ}+N\to \Lambda_c^+ +  D^-$ is already open for the $1P$ charmonia. This reaction may generate a non-negligible absorptive width for the $\chi_{cJ}$--nucleus bound states. For $\chi_{c0}(1P)$, this broadening is accompanied by its already sizable free-space natural width. Consequently, the $\chi_{cJ}$--nucleus bound states, if formed, should not be expected to appear as very narrow peaks in the same sense as the predicted $J/\psi$--nucleus bound states~\cite{Tsushima:2011kh}. The open-charm absorption channel may also provide an experimental handle on the $\chi_{cJ}$--$N$ interaction. In addition to this absorption process, radiative decay may offer a complementary route to searching for these states, for example through $\chi_{cJ}$--nucleus $\to$ $J/\psi$--nucleus $+\gamma$. Searches for such broadened \(\chi_{cJ}\)--nucleus structures, together with experimental tests of the predicted level ordering and level-spacing systematics, would therefore be well motivated in future high-statistics near-threshold photoproduction experiments at the upgraded JLab facility~\cite{Paryev:2026lzf}.

\section{Effective Cosh Potential for heavy-quarkonium--nucleus Bound States}

The systematic behavior discussed in Sec.~III indicates that the \(\chi_{cJ}\)--nucleus spectra are largely controlled by the radial shape of the effective potential. In the present framework, the numerical \(\chi_{cJ}\)--nucleus potentials are obtained by mapping the QMC density-dependent mass shifts onto finite nuclei within the local density approximation. These potentials will be referred to as the QMC-LDA potentials. Since their radial dependence mainly follows the finite nuclear density distribution, while their depths are fixed by the corresponding in-medium mass shifts, it is useful to examine whether they can be represented by a simple analytic density-like form.

For this purpose, we introduce the following cosh-type parametrization:
\begin{equation}
V^{(J,A,\alpha)}(r)=-V_0^{(J,\alpha)}\frac{1+\cosh(R_A/a_s)}{\cosh(r/a_s)+\cosh(R_A/a_s)} .
\label{eq:cosh_potential}
\end{equation}
Here, \(V_0^{(J,\alpha)}\) is the central depth, \(R_A=r_0 A^{1/3}\), and \(r_0\) and \(a_s\) are the effective radius and surface-diffuseness parameters, respectively. This parametrization corresponds to the normalized cosh form of a symmetrized Woods--Saxon potential and is closely related to the symmetrized Fermi function. Fermi-type distributions and Woods--Saxon potentials are widely used to describe finite-nucleus densities and phenomenological single-particle potentials~\cite{deVries1987,WoodsSaxon1954,Grypeos1991,Grypeos2001}. The cosh form is smooth at the origin, has a vanishing central slope, and retains a Woods--Saxon-like surface behavior. Woods--Saxon-type forms, including cosh and symmetrized versions, have been applied to cluster--core potentials and to nuclear, hypernuclear, and metal-cluster systems~\cite{BuckPilt1977,Grypeos2001}. They have also been employed in tunneling calculations of proton radioactivity~\cite{Chen2021}, cluster decay~\cite{Soylu2018}, and \(\alpha\) decay~\cite{BuckMerchantPerez1992}.

To test whether the QMC-LDA spectra can be described by a universal finite-nucleus geometry, the two geometry parameters are determined from the excitation-energy differences \(\Delta E_{n\ell}^{(J,A)}\), using the \(1p\), \(1d\), and \(2s\) levels. The absolute \(1s\) bound state energies are then used to fix the state- and cutoff-dependent depths \(V_0^{(J,\alpha)}\). The resulting common geometry is
\begin{equation}
r_0=1.05~{\rm fm},
\qquad
a_s=0.53~{\rm fm}.
\end{equation}
The fitted central depths are listed in Table~\ref{tab:cosh-depths-chicj}.

\begin{table}[htbp]
    \centering
    \caption{Fitted central depths \(V_0^{(J,\alpha)}\) of the cosh-type
    \(\chi_{cJ}\)--nucleus potential for different cutoff parameters
    \(\alpha\), in units of MeV.}
    \label{tab:cosh-depths-chicj}
    \begin{tabular*}{0.95\columnwidth}{@{\extracolsep{\fill}}cccc}
        \toprule\toprule
        State & \(\alpha=2\) & \(\alpha=3\) & \(\alpha=4\) \\
        \midrule
        \(\chi_{c0}(1P)\) & \(33.74\) & \(48.03\) & \(58.31\) \\
        \(\chi_{c1}(1P)\) & \(33.29\) & \(48.02\) & \(58.30\) \\
        \(\chi_{c2}(1P)\) & \(61.74\) & \(84.08\) & \(98.09\) \\
        \bottomrule\bottomrule
    \end{tabular*}
\end{table}

To quantify the agreement between the cosh parametrization and the reference calculations, we define the root-mean-square (RMS) deviation for an observable $X$ as
\begin{equation}
    \mathrm{RMS}(X)
    =
    \left[
    \frac{1}{N}
    \sum_{i=1}^{N}
    \left(
    X_i^{\mathrm{cosh}}
    -
    X_i^{\mathrm{ref}}
    \right)^2
    \right]^{1/2},
\end{equation}
where $X$ denotes either the bound state energy $E$ or the level
spacing $\Delta E$, $X_i^{\mathrm{cosh}}$ and $X_i^{\mathrm{ref}}$ are the corresponding values obtained from the cosh parametrization and the reference calculation, respectively, and $N$ is the total number of common levels included in the comparison.

With this definition, the cosh potential reproduces the QMC-LDA excitation spectra with sub-MeV accuracy, with a root-mean-square residual \({\rm RMS}(\Delta E)=0.638~{\rm MeV}\). For the absolute bound state energies, the residuals are larger but remain at the few-MeV level, \({\rm RMS}(E)=2.125~{\rm MeV}\). This indicates that the finite-nucleus radial dependence of the QMC-LDA potentials is well described by the common cosh geometry, whereas the dependence on the charmonium state and on the cutoff parameter is mainly carried by the central depth \(V_0^{(J,\alpha)}\).

The fitted central depths are numerically close to the magnitudes of the corresponding \(\chi_{cJ}(1P)\) mass shifts at normal nuclear matter density. This motivates a simple extension of the parametrization. Once the geometry parameters \(r_0\) and \(a_s\) are fixed, the bound-state spectrum of a heavy quarkonium state in a finite nucleus can be estimated by using its mass shift in nuclear matter at \(\rho_0\) as the input depth.

To test this prescription beyond the \(\chi_{cJ}(1P)\) sector, we apply the same cosh geometry to several heavy-quarkonium--nucleus systems for which bound-state spectra have been reported in the literature. For each selected system, the quoted mass shift at normal nuclear matter density is used as the input depth, and the corresponding finite-nucleus levels are calculated.

The  comparison is summarized in Table~\ref{tab:benchmark_summary}. The first four rows use spectra obtained in QMC-loop calculations as references~\cite{Tsushima:2011kh,Cobos-Martinez:2020ynh,Zeminiani:2021vaq,Kaur:2026qzf},\footnote{Here the QMC-loop approach denotes calculations in which the in-medium \(D^{(*)}\) or \(B^{(*)}\) meson masses are first obtained from the QMC model, and the heavy-quarkonium mass shifts are then evaluated from the corresponding \(D^{(*)}\bar D^{(*)}\) or \(B^{(*)}\bar B^{(*)}\) meson-loop contributions. This procedure is conceptually similar to that adopted in the present work. By contrast, the chiral SU(3)--QCDSR loop calculation quoted below obtains the in-medium \(D^{(*)}\) masses from the chiral SU(3) hadronic model combined with QCD sum rules, and then evaluates the \(J/\psi\) mass shift from the \(D^{(*)}\bar D^{(*)}\) loop contribution.} namely the \(J/\psi\) and \(\eta_c\) results with \(\Lambda_D=2000~{\rm MeV}\) and \(\Lambda_D=2500~{\rm MeV}\), respectively, and the \(\Upsilon\) and \(\eta_b\) results with \(\Lambda_B=4000~{\rm MeV}\). The last row corresponds to the chiral SU(3)--QCDSR loop calculation of the \(J/\psi\) spectrum with \(\kappa=0\) and \(\Lambda_D=3~{\rm GeV}\). A more detailed, level-by-level comparison is given in Appendix~\ref{B}, Table~\ref{tab:appendix_benchmark_unified}.

\begin{table}[t]
	\centering
	\caption{Summary of the benchmark comparison between the cosh potential and finite-nucleus reference calculations. The input depth $V_0$ denotes the magnitude of the central attractive potential, i.e. $V(0)= -V_0$. The RMS value denotes the root-mean-square deviation between the bound state energies obtained with the cosh potential and the corresponding reference energies over the common bound states. All energies are in MeV.}
	\label{tab:benchmark_summary}
	\setlength{\tabcolsep}{3.5pt}
	\renewcommand{\arraystretch}{1.18}
	\begin{tabular*}{\columnwidth}{@{\extracolsep{\fill}}cccc@{}}
		\toprule
        \toprule
		State
		& $V_0$
		& Nuclei
		& RMS \\
		\midrule
		$J/\psi$~\cite{Tsushima:2011kh}
		& $20.0$
		& \begin{tabular}[c]{@{}c@{}}
			$^{4}{\rm He}$, $^{12}{\rm C}$, $^{16}{\rm O}$ \\
			$^{40}{\rm Ca}$, $^{90}{\rm Zr}$, $^{208}{\rm Pb}$
		\end{tabular}
		& $2.15$ \\
		\addlinespace[0.45em]
		
		$\eta_c$~\cite{Cobos-Martinez:2020ynh}
		& $20.0$
		& \begin{tabular}[c]{@{}c@{}}
			$^{4}{\rm He}$, $^{12}{\rm C}$, $^{16}{\rm O}$, $^{40,48}{\rm Ca}$ \\
			$^{90}{\rm Zr}$, $^{197}{\rm Au}$, $^{208}{\rm Pb}$
		\end{tabular}
		& $2.61$ \\
		\addlinespace[0.45em]
		
		$\Upsilon$~\cite{Zeminiani:2021vaq}
		& $18.0$
		& \begin{tabular}[c]{@{}c@{}}
			$^{4}{\rm He}$, $^{12}{\rm C}$, $^{16}{\rm O}$, $^{40,48}{\rm Ca}$ \\
			$^{90}{\rm Zr}$, $^{197}{\rm Au}$, $^{208}{\rm Pb}$
		\end{tabular}
		& $1.88$ \\
		\addlinespace[0.45em]
		
		$\eta_b$~\cite{Zeminiani:2021vaq}
		& $78.0$
		& \begin{tabular}[c]{@{}c@{}}
			$^{4}{\rm He}$, $^{12}{\rm C}$, $^{16}{\rm O}$, $^{40,48}{\rm Ca}$ \\
			$^{90}{\rm Zr}$, $^{197}{\rm Au}$, $^{208}{\rm Pb}$
		\end{tabular}
		& $6.08$ \\
		\addlinespace[0.45em]
		
		$J/\psi$~\cite{Kaur:2026qzf}
		& $14.3$
		& \begin{tabular}[c]{@{}c@{}}
			$^{16}{\rm O}$, $^{40}{\rm Ca}$ \\
			$^{90}{\rm Zr}$, $^{208}{\rm Pb}$
		\end{tabular}
		& $1.14$ \\
		\bottomrule
        \bottomrule
	\end{tabular*}
\end{table}

As shown by the RMS values in Table~\ref{tab:benchmark_summary}, the absolute bound state energies obtained with the cosh potential are in good overall agreement with the reference calculations. Except for the deepest \(\eta_b\) potential, the RMS deviations for most systems are in the range of \(1\)--\(3~{\rm MeV}\). This is comparable to the accuracy obtained for the absolute \(\chi_{cJ}\) bound state energies in the fit to the QMC-LDA spectra, where \({\rm RMS}(E)=2.125~{\rm MeV}\). Since the benchmark calculations use only the quoted nuclear-matter mass shift as the input depth, without refitting the finite-nucleus spectra, this level of agreement indicates that the cosh parametrization provides a useful representation of the effective heavy-quarkonium--nucleus potential.

The larger RMS value obtained for the \(\eta_b\) case should be interpreted with some care. The corresponding input potential is substantially deeper, \(V_0\simeq 78~{\rm MeV}\), than those used for the \(J/\psi\), \(\eta_c\), and \(\Upsilon\) systems. For such a deep potential, the bound-state wave functions probe the nuclear interior more strongly, and small differences between the cosh profile and the reference potential can lead to larger absolute shifts in the eigenvalues. Although the deviations are larger in MeV, they remain roughly at the ten-percent level when compared with the much larger bound state energy scale for medium and heavy nuclei.

The calcium results provide an additional check of the interpolation capability of the parametrization. The agreement remains reasonable for calcium nuclei, including cases that were not used in fixing the potential parameters. This suggests that the cosh form not only reproduces a limited set of fitted reference values but also provides a stable description as the nuclear size and density profile are varied.

The largest discrepancies occur for some levels in the lightest nucleus, \({}^{4}\mathrm{He}\). This is not unexpected, since such a light system lies near the lower limit of applicability of a smooth mean-field potential and of the local density treatment used to construct an effective radial interaction. In light nuclei, finite-size effects and surface effects can become comparatively more important. Weakly bound or near-threshold states are also more sensitive to small changes in the potential shape. The deviations in \({}^{4}\mathrm{He}\) should therefore be viewed primarily as a limitation of the underlying smooth-potential description.

Moreover, the last row of Table~\ref{tab:benchmark_summary} shows that the same parametrization also gives a reasonable description of the \(J/\psi\) spectrum obtained in the chiral SU(3)--QCDSR loop calculation. The corresponding RMS deviation is \(1.14~{\rm MeV}\). This comparison suggests that the applicability of the cosh geometry is largely independent of the microscopic origin of the nuclear-matter mass shift, provided that the resulting finite-nucleus potential exhibits a density-like radial profile.

To examine whether the cosh parametrization preserves the level-spacing systematics discussed in Sec.~III, we compare the individual bound state energies with the corresponding reference results in Table~\ref{tab:appendix_benchmark_unified}. For medium and heavy nuclei, the deviations of different levels within a given spectrum are generally similar in magnitude. They therefore behave approximately as a common energy shift and largely cancel when level differences are considered. Consequently, the cosh potential reproduces not only the overall binding scale but also the level ordering and the characteristic spacing pattern of the reference finite-nucleus spectra. This cancellation is less effective in very light nuclei and for weakly bound states, whose energies are more sensitive to the surface shape and asymptotic tail of the potential.

Combining the comparisons of absolute bound state energies and level spacings, the results support the use of the cosh potential as a compact parametrization of the in-medium heavy-quarkonium--nucleus interaction. The parametrization is not intended to reproduce all microscopic details of the reference potentials. Rather, it provides a simple analytic representation that captures the leading effects of the potential depth, nuclear size, and surface profile on the finite-nucleus bound state spectrum.

The above conclusion should be understood within the assumptions underlying the unified parametrization. In practice, the cosh form is expected to be applicable when: (i) the local density approximation is valid, so that the finite-nucleus potential can be constructed from the density-dependent in-medium mass shift; (ii) the dominant heavy-quarkonium--nucleus interaction can be represented by a scalar potential; (iii) the quarkonium momentum is low; and (iv) the in-medium mass shift is approximately linear in the nuclear density, \(\Delta m(\rho)\propto\rho\), over the density range relevant for finite nuclei. Under these conditions, the density dependence of the microscopic input can be mapped onto a simple radial potential, and the cosh parametrization provides a compact representation of the resulting finite-nucleus spectra.

\section{DISCUSSIONS AND CONCLUSIONS}

In this work, we have investigated the possible formation of $\chi_{cJ}$--nuclear bound states in finite nuclei within a QMC-based framework. Since the $\chi_{cJ}(1P)$ states are hidden-charm $c\bar c$ mesons, they do not couple directly to the scalar and vector mean fields at the valence quark level. Their in-medium mass shifts are generated indirectly through the density dependence of virtual $D^{(*)}\bar D^{(*)}$ loop contributions in an unquenched picture~\cite{Tsushima:2011kh,Zeminiani:2020aho,Zeminiani:2023gqc,Zhang:2025fol}. The resulting nuclear-matter mass shifts are then converted into radial $\chi_{cJ}$--nucleus potentials through the local density approximation, and the bound state energies are obtained by solving a Klein--Gordon-type equation~\cite{Tsushima:2011kh}.

The calculated spectra show that the attractive $\chi_{cJ}$--nucleus potentials support bound states for all nuclei considered in this work, namely $^{12}{\rm C}$, $^{16}{\rm O}$, $^{40}{\rm Ca}$, $^{90}{\rm Zr}$, $^{197}{\rm Au}$, and $^{208}{\rm Pb}$. The lowest $1s$ bound state energies of a given $\chi_{cJ}(1P)$ state become nearly degenerate among heavy nuclei. For a fixed nucleus, the $\chi_{c0}(1P)$ and $\chi_{c1}(1P)$ spectra are also nearly degenerate because their in-medium mass shifts are very similar, whereas the larger downward mass shift of $\chi_{c2}(1P)$ produces substantially deeper $\chi_{c2}$--nucleus states. Increasing the cutoff parameter $\alpha$ mainly deepens the potential and shifts the entire spectrum downward; in medium and heavy nuclei, the lowest $\chi_{c2}$ levels can reach bound state energies of approximately $-100$ MeV for $\alpha=4$. The relative level structure, however, exhibits a considerably weaker dependence on $\alpha$. In particular, the energy spacings are well described by the scaling relation $\Delta E_{n\ell}^{(J,A)} \propto N_{n\ell}A^{-2/3}$, with $N_{n\ell}=2(n-1)+\ell$.

Motivated by these spectral systematics, in Sec.~IV we introduced a cosh-type potential as a compact analytic representation of the QMC-LDA potentials. A common finite-nucleus geometry, $r_0=1.05~{\rm fm}$ and $a_s=0.53~{\rm fm}$, reproduces the $\chi_{cJ}$ excitation energies with ${\rm RMS}(\Delta E)=0.638~{\rm MeV}$ and their absolute bound state energies with ${\rm RMS}(E)=2.125~{\rm MeV}$, while the dependence on the charmonium state and the cutoff parameter is absorbed primarily into the central depth. Using the nuclear-matter mass shift at $\rho_0$ as the input depth, the same geometry also gives a reasonable description of previously reported $J/\psi$, $\eta_c$, $\Upsilon$, and $\eta_b$ nuclear spectra obtained from different microscopic calculations. The agreement is generally at the few-MeV level, with larger deviations occurring for very light nuclei and for particularly deep potentials. The cosh parametrization should therefore be regarded as a useful compact representation of heavy-quarkonium--nucleus potentials, and its applicability relies on the validity of the local density approximation, a predominantly scalar interaction at low momentum, and an approximately linear density dependence of the in-medium mass shift.

The results obtained here provide a first systematic estimate of $\chi_{cJ}$--nuclear bound state spectra based on the QMC model and the unquenched description of in-medium $P$-wave charmonia. They suggest that finite nuclei can support bound $\chi_{cJ}(1P)$ states over a broad range of nuclear masses and cutoff parameters. It would be worthwhile in future studies to include the imaginary part of the optical potential, evaluate formation spectra for realistic experimental reactions, and examine the combined effects of finite density and finite temperature. Such investigations will be useful for assessing the observability of $\chi_{cJ}$--nuclear states at future high-statistics facilities and for clarifying the properties of charmonium in cold nuclear matter.

\begin{acknowledgments}
We would like to thank Prof. Anthony Thomas for useful discussion. This work is supported by the Natural Science Foundation of Gansu Province (No. 26RCKA012 and No. 25JRRA799), the National Natural Science Foundation of China under Grants No. 12335001 and No. 12247101, the ``111 Center" under Grant No. B20063, the fundamental Research Funds for the Central Universities (lzujbky-2023-stlt01), and Lanzhou City High-Level Talent Funding. T.-L. G. is supported by the Gansu Province Postgraduate Innovation Star Program No. 2026CXZX-042.

\end{acknowledgments}

\appendix

\section{Bound state energies}\label{A}
In this appendix, we list all numerical values of the bound state energies in Table~\ref{tab:chic-binding}. 
\begin{table*}[t]
    \centering
    \caption{Bound state energies $E^{(J,\,A)}_{n\ell}$ in MeV for $\chi_{cJ}(1P)$ states.}
    \label{tab:chic-binding}

    \small
    \renewcommand{\arraystretch}{1.15}
    \setlength{\tabcolsep}{5.5pt}

    \begin{adjustbox}{max width=\textwidth}
        \begin{tabular}{
            @{}
            l
            l
            *{9}{S[table-format=-2.2]}
            @{}
        }
            \toprule\toprule
            &
				& \multicolumn{3}{c}{$\chi_{c0}(1P)$}
				& \multicolumn{3}{c}{$\chi_{c1}(1P)$}
				& \multicolumn{3}{c}{$\chi_{c2}(1P)$}
				\\
				\cmidrule(lr){3-5}
				\cmidrule(lr){6-8}
				\cmidrule(lr){9-11}
				&
				& \multicolumn{1}{c}{$\alpha=2$}
				& \multicolumn{1}{c}{$\alpha=3$}
				& \multicolumn{1}{c}{$\alpha=4$}
				& \multicolumn{1}{c}{$\alpha=2$}
				& \multicolumn{1}{c}{$\alpha=3$}
				& \multicolumn{1}{c}{$\alpha=4$}
				& \multicolumn{1}{c}{$\alpha=2$}
				& \multicolumn{1}{c}{$\alpha=3$}
				& \multicolumn{1}{c}{$\alpha=4$}
				\\
				\midrule
				
				${}^{12}_{\chi_{cJ}}\mathrm{C}$   & $1s$ &  -18.56 & -30.84 & -39.78 & -18.50 & -30.56 & -39.27 & -42.30 & -61.83 & -74.31 \\
				& $1p$ &  -9.24 &  -19.96 &  -27.91 & -9.30 & -19.36 & -26.95 & -29.64 & -47.18 & -58.53 \\
				& $1d$ &  -1.05 &  -9.45 &  -16.19 & -1.17 & -8.65 & -14.89 & -17.19 & -32.45 & -42.56 \\
				& $2s$ &  -2.34 &  -9.93 &  -16.36 & -2.42 & -9.29 & -15.22 & -17.42 & -32.23 & -42.14 \\
				
				${}^{16}_{\chi_{cJ}}\mathrm{O}$   & $1s$ & -22.04 & -34.55 & -43.78 & -21.95 & -34.91 & -44.17 & -47.37 & -67.92 & -80.97 \\
				& $1p$ &  -13.82 &  -25.09 &  -33.63 & -13.87 & -25.53 & -34.06 & -37.08 & -56.39 & -68.77 \\
				& $1d$ &  -5.51 &  -15.09 &  -22.71 & -5.69 & -15.59 & -23.19 & -25.96 & -43.71 & -55.25 \\
				& $2s$ &  -5.43 &  -14.21 &  -21.39 & -5.57 & -14.67 & -21.86 & -24.50 & -41.59 & -52.78 \\
				
				${}^{40}_{\chi_{cJ}}\mathrm{Ca}$  & $1s$ & -29.80 & -44.06 & -54.46 & -29.19 & -44.01 & -54.48 & -58.14 & -81.24 & -95.92 \\
				& $1p$ & -23.97 & -37.42 & -47.30 & -23.50 & -37.48 & -47.43 & -50.93 & -73.00 & -87.05 \\
				& $1d$ & -17.95 & -30.60 & -40.00 & -17.62 & -30.77 & -40.23 & -43.61 & -64.78 & -78.30 \\
				& $2s$ & -17.64 & -30.31 & -39.76 & -17.30 & -30.48 & -40.02 & -43.44 & -64.82 & -78.48 \\
				
				${}^{90}_{\chi_{cJ}}\mathrm{Zr}$  & $1s$ & -32.13 & -46.64 & -57.18 & -31.85 & -46.94 & -57.54 & -61.09 & -84.33 & -99.15 \\
				& $1p$ & -28.76 & -42.95 & -53.28 & -28.56 & -43.30 & -53.69 & -57.07 & -79.93 & -94.69 \\
				& $1d$ & -24.95 & -38.76 & -48.85 & -24.84 & -39.18 & -49.33 & -52.50 & -74.89 & -89.54 \\
				& $2s$ & -24.13 & -37.82 & -47.83 & -24.03 & -38.25 & -48.32 & -51.44 & -73.66 & -88.21 \\
				
				${}^{197}_{\chi_{cJ}}\mathrm{Au}$  & $1s$ & -31.09 & -44.85 & -54.83 & -30.83 & -45.14 & -55.17 & -58.53 & -80.54 & -94.37 \\
				& $1p$ & -29.59 & -43.31 & -53.26 & -29.38 & -43.63 & -53.64 & -56.90 & -78.97 & -92.79 \\
				& $1d$ & -27.65 & -41.26 & -51.15 & -27.48 & -41.62 & -51.57 & -54.70 & -76.81 & -90.57 \\
				& $2s$ & -26.50 & -39.93 & -49.70 & -26.36 & -40.31 & -50.13 & -53.17 & -75.07 & -88.67 \\
				
				${}^{208}_{\chi_{cJ}}\mathrm{Pb}$ & $1s$ & -32.56 & -46.73 & -56.99 & -31.88 & -46.60 & -56.93 & -60.39 & -83.07 & -97.33 \\
				& $1p$ & -30.68 & -44.69 & -54.84 & -30.06 & -44.60 & -54.82 & -58.14 & -80.72 & -94.86 \\
				& $1d$ & -28.54 & -42.37 & -52.42 & -27.97 & -42.34 & -52.45 & -55.63 & -78.14 & -92.16 \\
				& $2s$ & -28.03 & -41.87 & -51.94 & -27.47 & -41.85 & -51.98 & -55.12 & -77.73 & -91.78 \\
				
            \bottomrule\bottomrule
        \end{tabular}
    \end{adjustbox}
\end{table*}

\section{Level-by-level comparison of the cosh parametrization}\label{B}
In this appendix, we present the detailed level-by-level comparison between the bound state energies obtained from the cosh parametrization and those from the corresponding reference calculations in Table~\ref{tab:appendix_benchmark_unified}. The common geometry parameters \(r_0=1.05~{\rm fm}\) and \(a_s=0.53~{\rm fm}\) are used throughout, while for each quarkonium state the potential depth \(V_0\) is determined from the magnitude of the mass shift at normal nuclear-matter density, \(\rho=\rho_0\), quoted in the corresponding reference calculation.
\begin{table*}[!t]
	\centering
	\caption{Detailed level-by-level comparison of the bound state energies obtained with the cosh parametrization and the corresponding reference calculations. Here, $C$ and $R$ denote the energies obtained from the cosh parametrization and the reference calculations, respectively, and $\Delta=C-R$. The symbol -- indicates that the corresponding nucleus--state combination was not considered in the reference calculation, whereas $\times$ denotes that no bound state was obtained for the corresponding channel. All energies are given in MeV.}
	\label{tab:appendix_benchmark_unified}
	\setlength{\tabcolsep}{2.0pt}
	\renewcommand{\arraystretch}{0.88}
	\providecommand{\tabdash}{\multicolumn{1}{c}{--}}
	\begin{adjustbox}{max width=\textwidth,max totalheight=0.88\textheight,keepaspectratio}
		\begin{tabular}{@{}lc*{15}{>{$}c<{$}}@{}}
			\toprule
            \toprule
			& &
			\multicolumn{3}{c}{$J/\psi$~\cite{Tsushima:2011kh}} &
			\multicolumn{3}{c}{$\eta_c$~\cite{Cobos-Martinez:2020ynh}} &
			\multicolumn{3}{c}{$\Upsilon$~\cite{Zeminiani:2021vaq}} &
			\multicolumn{3}{c}{$\eta_b$~\cite{Zeminiani:2021vaq}} &
			\multicolumn{3}{c}{$J/\psi$~\cite{Kaur:2026qzf}} \\
			\cmidrule(lr){3-5}
			\cmidrule(lr){6-8}
			\cmidrule(lr){9-11}
			\cmidrule(lr){12-14}
			\cmidrule(l){15-17}
			Nucleus & State
			& C & R & \Delta
			& C & R & \Delta
			& C & R & \Delta
			& C & R & \Delta
			& C & R & \Delta \\
			\midrule
			
			$^{4}\mathrm{He}$ & $1s$
			& -3.62 & -5.74 & 2.12
			& -3.51 & -5.49 & 1.98
			& -5.08 & -7.5 & 2.42
			& -49.88 & -66.7 & 16.82
			& \tabdash & \tabdash & \tabdash \\
			& $1p$
			& \times & \times & \times
			& \times & \times & \times
			& \times & \times & \times
			& -30.75 & -43.7 & 12.95
			& \tabdash & \tabdash & \tabdash \\
			& $1d$
			& \times & \times & \times
			& \times & \times & \times
			& \times & \times & \times
			& -12.99 & -19.7 & 6.71
			& \tabdash & \tabdash & \tabdash \\
			& $2s$
			& \times & \times & \times
			& \times & \times & \times
			& \times & \times & \times
			& -14.63 & -17.9 & 3.27
			& \tabdash & \tabdash & \tabdash \\
			\addlinespace[0.2em]
			
			$^{12}\mathrm{C}$ & $1s$
			& -9.70 & -11.21 & 1.51
			& -9.54 & -11.28 & 1.74
			& -11.43 & -12.8 & 1.37
			& -65.89 & -69.0 & 3.11
			& \tabdash & \tabdash & \tabdash \\
			& $1p$
			& -2.70 & -3.94 & 1.24
			& -2.48 & -3.69 & 1.21
			& -6.60 & -7.9 & 1.30
			& -56.37 & -60.1 & 3.73
			& \tabdash & \tabdash & \tabdash \\
			& $1d$
			& \times & \times & \times
			& \times & \times & \times
			& -1.90 & -2.9 & 1.00
			& -46.29 & -50.4 & 4.11
			& \tabdash & \tabdash & \tabdash \\
			& $2s$
			& \times & \times & \times
			& \times & \times & \times
			& -2.10 & -2.8 & 0.70
			& -45.35 & -49.1 & 3.75
			& \tabdash & \tabdash & \tabdash \\
			\addlinespace[0.2em]
			
			$^{16}\mathrm{O}$ & $1s$
			& -11.10 & -13.26 & 2.16
			& -10.95 & -13.15 & 2.20
			& -12.64 & -14.2 & 1.56
			& -68.49 & -71.0 & 2.51
			& -6.65 & -7.827 & 1.18 \\
			& $1p$
			& -4.58 & -6.81 & 2.23
			& -4.34 & -6.48 & 2.14
			& -8.50 & -10.4 & 1.90
			& -60.77 & -64.9 & 4.13
			& -1.41 & -2.133 & 0.72 \\
			& $1d$
			& \times & \times & \times
			& \times & \times & \times
			& -4.23 & -6.2 & 1.97
			& -52.45 & -57.9 & 5.45
			& \tabdash & \tabdash & \tabdash \\
			& $2s$
			& \times & \times & \times
			& \times & \times & \times
			& -4.04 & -5.4 & 1.36
			& -51.40 & -56.3 & 4.90
			& \tabdash & \tabdash & \tabdash \\
			\addlinespace[0.2em]
			
			$^{40}\mathrm{Ca}$ & $1s$
			& -14.73 & -17.24 & 2.51
			& -14.61 & -18.31 & 3.70
			& -15.31 & -18.2 & 2.89
			& -73.84 & -82.6 & 8.76
			& -9.59 & -8.563 & -1.03 \\
			& $1p$
			& -10.19 & -12.92 & 2.73
			& -9.97 & -13.59 & 3.62
			& -12.97 & -15.9 & 2.93
			& -70.01 & -79.0 & 8.99
			& -5.64 & -5.054 & -0.59 \\
			& $1d$
			& -5.24 & -8.21 & 2.97
			& -4.93 & -8.36 & 3.43
			& -10.33 & -13.3 & 2.97
			& -65.70 & -74.9 & 9.20
			& -1.49 & -1.903 & 0.41 \\
			& $2s$
			& -4.53 & -7.48 & 2.95
			& -4.24 & -7.44 & 3.20
			& -9.77 & -12.7 & 2.93
			& -64.77 & -74.0 & 9.23
			& -1.25 & -2.335 & 1.08 \\
			\addlinespace[0.2em]
			
			$^{48}\mathrm{Ca}$ & $1s$
			& \tabdash & \tabdash & \tabdash
			& -15.19 & -18.26 & 3.07
			& -15.66 & -17.9 & 2.24
			& -74.51 & -80.2 & 5.69
			& \tabdash & \tabdash & \tabdash \\
			& $1p$
			& \tabdash & \tabdash & \tabdash
			& -10.94 & -14.23 & 3.29
			& -13.59 & -16.0 & 2.41
			& -71.20 & -77.4 & 6.20
			& \tabdash & \tabdash & \tabdash \\
			& $1d$
			& \tabdash & \tabdash & \tabdash
			& -6.23 & -9.63 & 3.40
			& -11.23 & -13.8 & 2.57
			& -67.43 & -74.2 & 6.77
			& \tabdash & \tabdash & \tabdash \\
			& $2s$
			& \tabdash & \tabdash & \tabdash
			& -5.39 & -8.54 & 3.15
			& -10.67 & -13.2 & 2.53
			& -66.56 & -73.3 & 6.74
			& \tabdash & \tabdash & \tabdash \\
			\addlinespace[0.2em]
			
			$^{90}\mathrm{Zr}$ & $1s$
			& -16.88 & -18.69 & 1.82
			& -16.79 & -19.14 & 2.35
			& -16.57 & -18.1 & 1.53
			& -76.15 & -78.9 & 2.75
			& -11.42 & -9.982 & -1.43 \\
			& $1p$
			& -13.91 & -16.07 & 2.16
			& -13.76 & -16.53 & 2.77
			& -15.23 & -17.0 & 1.77
			& -74.16 & -77.5 & 3.34
			& -8.74 & -7.306 & -1.44 \\
			& $1d$
			& -10.47 & -13.06 & 2.59
			& -10.22 & -13.38 & 3.16
			& -13.66 & -15.7 & 2.04
			& -71.83 & -75.7 & 3.87
			& -5.66 & -4.773 & -0.89 \\
			& $2s$
			& -9.49 & -12.22 & 2.73
			& -9.23 & -12.29 & 3.06
			& -13.17 & -15.2 & 2.03
			& -71.15 & -74.9 & 3.75
			& -4.87 & -4.963 & 0.09 \\
			\addlinespace[0.2em]
			
			$^{197}\mathrm{Au}$ & $1s$
			& \tabdash & \tabdash & \tabdash
			& -18.13 & -19.26 & 1.13
			& -17.23 & -17.7 & 0.47
			& -77.25 & -76.1 & -1.15
			& \tabdash & \tabdash & \tabdash \\
			& $1p$
			& \tabdash & \tabdash & \tabdash
			& -16.22 & -17.77 & 1.55
			& -16.47 & -17.2 & 0.73
			& -76.20 & -75.6 & -0.60
			& \tabdash & \tabdash & \tabdash \\
			& $1d$
			& \tabdash & \tabdash & \tabdash
			& -13.93 & -15.87 & 1.94
			& -15.54 & -16.4 & 0.86
			& -74.94 & -74.6 & -0.34
			& \tabdash & \tabdash & \tabdash \\
			& $2s$
			& \tabdash & \tabdash & \tabdash
			& -13.10 & -15.04 & 1.94
			& -15.20 & -16.0 & 0.80
			& -74.48 & -74.0 & -0.48
			& \tabdash & \tabdash & \tabdash \\
			\addlinespace[0.2em]
			
			$^{208}\mathrm{Pb}$ & $1s$
			& -18.25 & -19.10 & 0.85
			& -18.20 & -19.82 & 1.62
			& -17.27 & -18.2 & 0.93
			& -77.30 & -78.1 & 0.80
			& -12.62 & -11.146 & -1.48 \\
			& $1p$
			& -16.46 & -17.59 & 1.13
			& -16.36 & -18.37 & 2.01
			& -16.53 & -17.7 & 1.17
			& -76.30 & -77.5 & 1.20
			& -10.97 & -9.231 & -1.74 \\
			& $1d$
			& -14.31 & -15.81 & 1.50
			& -14.14 & -16.49 & 2.35
			& -15.64 & -16.9 & 1.26
			& -75.09 & -76.6 & 1.51
			& -8.98 & -7.367 & -1.61 \\
			& $2s$
			& -13.51 & -15.26 & 1.75
			& -13.32 & -15.70 & 2.38
			& -15.30 & -16.6 & 1.30
			& -74.65 & -76.0 & 1.35
			& -8.26 & -7.439 & -0.82 \\
			\bottomrule
            \bottomrule
		\end{tabular}
	\end{adjustbox}
\end{table*}

\bibliography{chicj_cleaned}

\end{document}